\documentclass[aps, reprint, longbibliography]{revtex4-2}[pra]
\usepackage{graphicx, natbib, amsmath, mhchem}
\usepackage{subfigure, amssymb}
\usepackage{float} 
\usepackage{color} 
\DeclareGraphicsExtensions{.pdf, .png}
\graphicspath{ {Figures/} }

\begin{document}

\title{Action potentials in vitro: theory and experiment}
\author{Ziqi Pi}
\author{Giovanni Zocchi}
\email{zocchi@physics.ucla.edu}
\affiliation{Department of Physics and Astronomy, University of California - Los Angeles}

\begin{abstract}
\noindent Action potential generation underlies some of the most consequential dynamical systems on Earth, from brains to hearts. It is therefore interesting to develop synthetic cell-free systems, based on the same molecular mechanisms, which may allow for the exploration of parameter regions and phenomena not attainable, or not apparent, in the live cell. We previously constructed such a synthetic system, based on biological components, which fires action potentials. We call it ``Artificial Axon''. The system is minimal in that it relies on a single ion channel species for its dynamics. Here we characterize the Artificial Axon as a dynamical system in time, using a simplified Hodgkin-Huxley model adapted to our experimental context. We construct a phase diagram in parameter space identifying regions corresponding to different 
temporal behavior, such as Action Potential (AP) trains, single shot APs, or damped oscillations. The main new result is the finding that our system with a single ion channel species, with inactivation, is dynamically equivalent to the system of two channel species without inactivation (the Morris-Lecar system), which exists in nature. We discuss the transitions and bifurcations occurring crossing phase boundaries in the phase diagram, and obtain criteria for the channels' properties necessary to obtain the desired dynamical behavior. \\ 
In the second part of the paper we present new experimental results obtained with a system of two AAs connected by excitatory and/or inhibitory electronic ``synapses''. We discuss the feasibility of constructing an autonomous oscillator 
with this system. 
\end{abstract}

\maketitle


\section {Introduction}
 
\noindent The physics of excitable media is largely concerned with the patterns in space and time created by nonlinear excitations. In this context, electrophysiological processes, and action potentials specifically, hold a unique place as some of the most consequential dynamical systems on earth. Action potentials (APs) play a vital role in biological computation, as sequences of APs encode information in a variety of ways \cite{gerstner1997}.  \\
We have recently introduced a minimal synthetic system, the "Artificial Axon" (AA) \cite{Amila2016, Hector2017, Ziqi2021}, which is capable of generating APs in time. The experimental system is based on a traditional suspended lipid bilayer (``black lipid membrane'') with embedded voltage gated potassium ion channels (KvAP). An ionic gradient maintained 
across the membrane provides the free energy source to elicit action potentials. Non-traditionally, the system is held 
in the off-equilibrium, excitable state by a modified voltage clamp (``Current Limited Voltage Clamp'': CLVC) 
which allows for voltage dynamics. The system is minimal in that it is built with one ion channel species only, yet it can support APs. The key is the dynamics of the channel, which includes inactivation. Previously we reported on the dynamics of firing APs in the AA. We showed that the threshold behavior of the system is the same as in real neurons 
\cite{Sejnowski2008}, namely it is governed by a saddle node bifurcation \cite{Ziqi2021}.

\noindent Here we first discuss numerically the phase diagram for a single AA, identifying regions in parameter space which give rise to trains of APs, damped oscillations, or single shot APs. We propose a simplified version of the Hodgkin-Huxley model \cite{hodgkin_quantitative_1952} for our system, based on the kinetics of our voltage gated ion channel KvAP \cite{MacKinnon_gatingModel2009}. The main new insight is that a system with one ion channel species, with inactivation, is dynamically equivalent to the (biological) system with two channel species, without inactivation (the Morris-Lecar system \cite{Morris_Lecar}).  We identify the bifurcations separating different regions in the phase diagram of the AA; in particular, we point out a transition which may not have been described before in elctrophysiology. In the second part of the paper we present experimental measurements of the inactivation and recovery rates of the channels in the AA system. These help to qualitatively place the present experimental system in the phase diagram obtained from the model, and understand the requirements on channel dynamics in order to obtain autonomous oscillations in the AA. Finally, we demonstrate a system of two AAs connected by electronic ``synapses'', as a prototype for future network developments.

\section{Theory} 

\subsection{The Artificial Axon System}

\noindent In the Artificial Axon, the phospholipid bilayer acts in essence as the dielectric of a parallel plates capacitance, sandwiched between two conducting media which are the electrolytes on either side. This capacitance is charged by two kinds of ionic currents: the current through the ion channels embedded in the membrane, and the current sourced by the clamp electrodes. The charge carriers for the former are K$^+$ ions, for the latter Cl$^-$, Ag$^+$, and all other ions in solution. The voltage dynamics $V(t)$ is governed by the following equation \cite{Ziqi2021}: 

\begin{equation}
\begin{split}
\frac{d V}{d t} = \frac{N_0 \chi}{C} \,  p_o(t) \,  [V_N - V(t)] 
+  \frac{\chi_c}{C} \, [V_{c} - V(t)] 
\end{split}
\label{eq:membrane}
\end{equation}

\noindent where $C$ is the membrane capacitance, $N_0$ the number of ion channels, $\chi$ the open channel conductance, $p_o(t)$ the fraction of channels in the open (conducting) state (so $N_0 \chi \, p_o(t)$ is the total channel conductance); $V_N$ is the Nernst potential corresponding to the bulk concentrations of $K^+$ ions on the two sides of the membrane, $\chi_c$ is the CLVC conductance, and $V_{c}$ the CLVC command voltage (which is a control parameter in the experiments). The first term on the RHS of (\ref{eq:membrane}) is the channel current (divided by $C$), proportional to the driving force $[V_N - V(t)]$, since $V_N$ is the equilibrium potential and $V(t)$ the present potential. The second term is the clamp current, proportional to the driving force $[V_{c} - V(t)]$. This second term is exactly equivalent to the presence of 
a second, reversed ionic gradient (of sodium ions, say) with Nernst potential equal to $V_{c}$ and corresponding leak conductance $\chi_c$ \cite{Amila2016, Ziqi2021}. Equations of the basic form (\ref{eq:membrane}) underlie many models of nerve excitability \cite{Koch_Book}.  

\begin{figure}
	\includegraphics[width=1.5in]{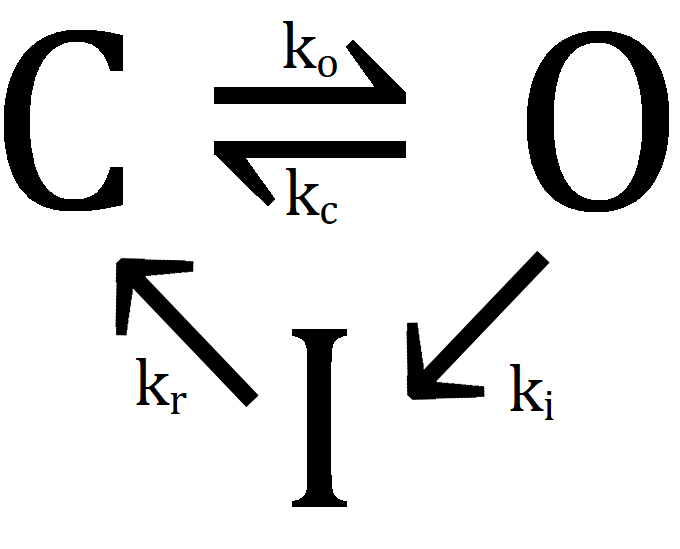}
	\caption{Simplified model for the KvAP channel dynamics. The three states are open (O), closed (C), and inactive (I).}
	\label{fig:model}
\end{figure}

The probability that channels are open, $p_o(t)$, is determined by the voltage dependent channel dynamics. The minimal model for the KvAP channel has 3 states: open, closed, and inactive (Fig.~\ref{fig:model}). In order to minimize the dimensions of parameter space, we connect them with 4 rate constants as shown in the figure. The unidirectional arrows are, strictly speaking, unphysical, but since here we are concerned with the macroscopic dynamics of the AA rather than the microscopic channel dynamics, they represent a permissible approximation. In fact, the detailed channel dynamics is more complex than shown in Fig.~\ref{fig:model}, with more states and corresponding transition rates \cite{MacKinnon_gatingModel2009}. We assume that these complications do not change the qualitative features of the phase diagram of the system, nor the nature of the bifurcations which occur in the dynamics. The rate equations which determine $p_o(t)$ in (\ref{eq:membrane}) are then: 

{\renewcommand{\arraystretch}{2.5}
\begin{normalsize}
\begin{equation}
\begin{aligned}
\left\{\begin{array}{ll}
 \cfrac{d p_i}{d t} = p_o(t) \, k_i(V) - p_i(t) \, k_r(V) \\
 \cfrac{d p_o}{d t} = (1 - p_o - p_i) \, k_o - p_o \, (k_c + k_i) 
\end{array}\right.
\end{aligned}
\label{eq:rates_1}
\end{equation}
\end{normalsize}
}

\noindent where $p_i(t)$ is the probability that the channels are in the inactive state $(p_o + p_i + p_c = 1)$, and the rates are as in Fig.~\ref{fig:model}. Since the channel is voltage gated, $k_o$ and $k_c$ are voltage dependent, which couples eqs.~(\ref{eq:rates_1}) to eq.~(\ref{eq:membrane}). We assume a standard Arrhenius dependence \cite{MacKinnon_gatingModel2009,ruta_functional_2003}: 
 
\begin{equation}
k_o(V) = \kappa \, e^{\alpha (V - V_0)} \quad , \quad k_c(V) = \kappa \, e^{- \alpha (V - V_0)}
\label{eq:rates_2}
\end{equation} 

\noindent this symmetric form being chosen once again to minimize the number of parameters \cite{Ziqi2021, Morris_Lecar}. The inactivation and recovery rates, $k_i$ and $k_r$, are similarly voltage dependent; however for a qualitative discussion of the phase diagram of the AA we may take them as constant. In the following, we refer to the model (\ref{eq:membrane}), (\ref{eq:rates_1}) with voltage independent $k_i$, $k_r$ as the ``voltage independent model''. The dynamical system (\ref{eq:membrane}), (\ref{eq:rates_1}) describes the dynamics of the physical AA well \cite{Ziqi2021}. For example, Fig.~\ref{fig:Push_Pull_sim_V} shows the time traces obtained numerically for two coupled AAs represented by (\ref{eq:membrane}), (\ref{eq:rates_1}), to be compared to the experimental traces in Fig. \ref{fig:Push_Pull_exp}. The parameter values (section III) used were obtained through a combination of previous measurements and fitting the experimental traces \cite{Hector2017, Ziqi2021}. \\ 

In order to understand the requirements on channel dynamics to obtain various temporal patterns, 
we construct a phase diagram of the dynamical behavior of the system (\ref{eq:membrane}), (\ref{eq:rates_1}), representing a single AA. The difficulty is that even in this minimal representation, the parameter space is still high dimensional. To make progress, we identify the most relevant parameters in relation to the experiments: the clamp voltage $V_{c}$ and clamp conductance $\chi_c$, which are the actual control parameter in the experiments, and the effective rates of Fig.~\ref{fig:model}, which define the suitability of the channel for obtaining interesting dynamical behavior, such as autonomous firing. In general (and specifically for the KvAP), the rate of opening and closing are much faster than those of inactivation and recovery, $k_o, k_c \gg k_i, k_r$, and so the most relevant parameters are then the clamp voltage, the clamp conductance \cite{Xinyi2022}, and the rates of inactivation and recovery. We may also cast the problem in dimensionless variables: eq.~(\ref{eq:membrane}) suggests the choice of $\tau = C /  (N_0 \chi)$ as the time scale and $V_N$ as the voltage scale; then the dimensionless control parameters in (\ref{eq:membrane}) 
are the clamp voltage $\tilde{V_c} = V_{c} / V_N$ and clamp conductance $\tilde{\chi_c} = \chi_c / (N_0 \chi)$, while the dimensionless rates are $\tilde{k_i} = \tau \, k_i = k_i \, C / (N_0 \chi)$ and similarly for $\tilde{k}_r$. In the following sections, we will begin our analysis with a dimensional 3D model before further reducing to a dimensionless 2D model.

\subsection{Phase Diagram} 

\begin{figure}
	\includegraphics[width=\linewidth]{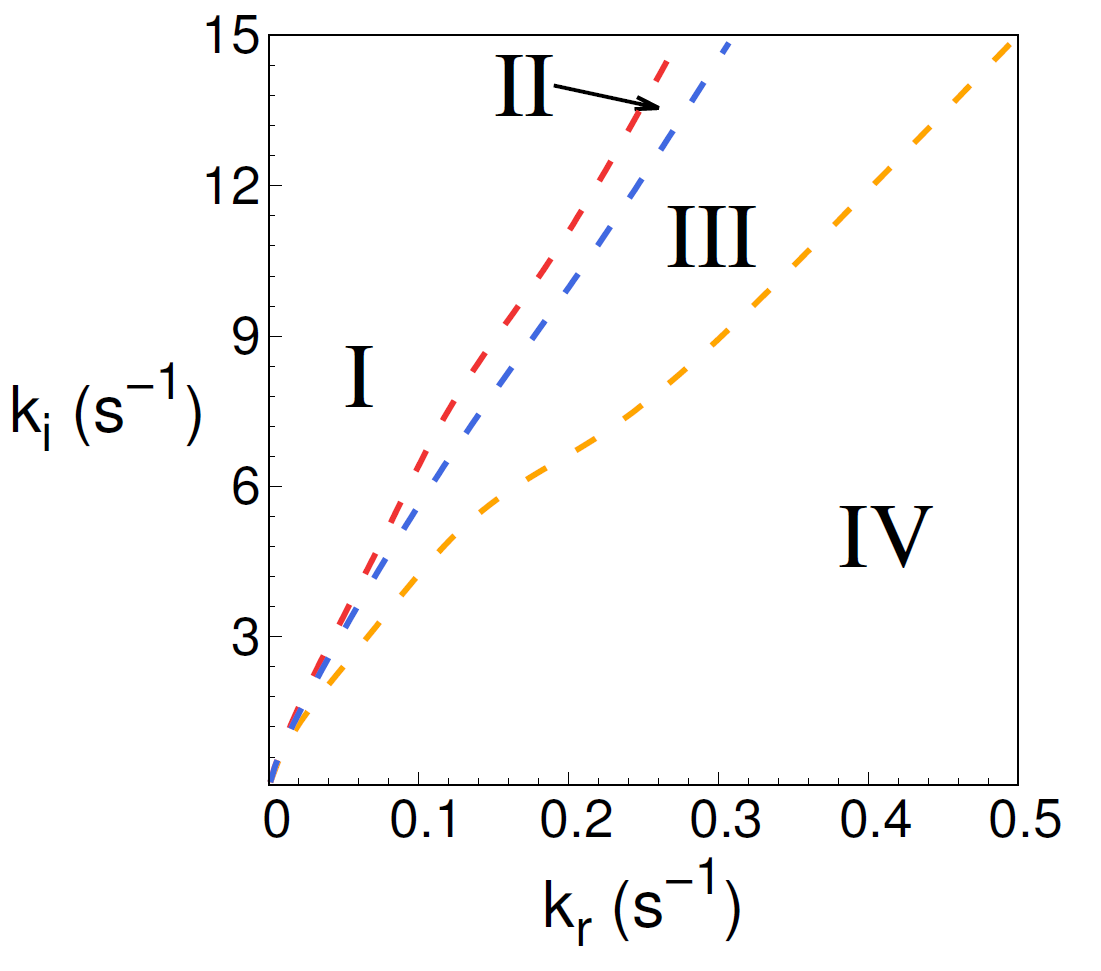}
	\caption{Phase diagram of the dynamic behavior obtained from the model (\ref{eq:membrane}), (\ref{eq:rates_1}), with voltage independent inactivation and recovery rates $k_i$ and $k_r$. (Fig.~\ref{fig:model}). The figure is a cut through a higher dimensional parameter space, at constant $\chi_c = 500 \,$pS and $V_c = - 50 \,$mV. Region I corresponds to AP trains. Region II exhibits smaller amplitude ``oscillations'', distinct from APs. Region III corresponds to damped oscillations, and Region IV to single shot APs.}
	\label{fig:pspace}
\end{figure}

Fig.~\ref{fig:pspace} shows one representation of the phase diagram (in dimensional variables) for the system (\ref{eq:membrane}), (\ref{eq:rates_1}) representing the AA, namely a cut through parameter space for constant  $\chi_c = 5 \times 10^{-10} \, \Omega^{-1} = 500 \,$pS and $V_{c} = -50 \,$mV, with $k_i$ and $k_r$ voltage independent. The diagram is generated by simluating AAs with the given parameters and recording the resulting time traces. \\

\begin{figure}
	\includegraphics[width=\linewidth]{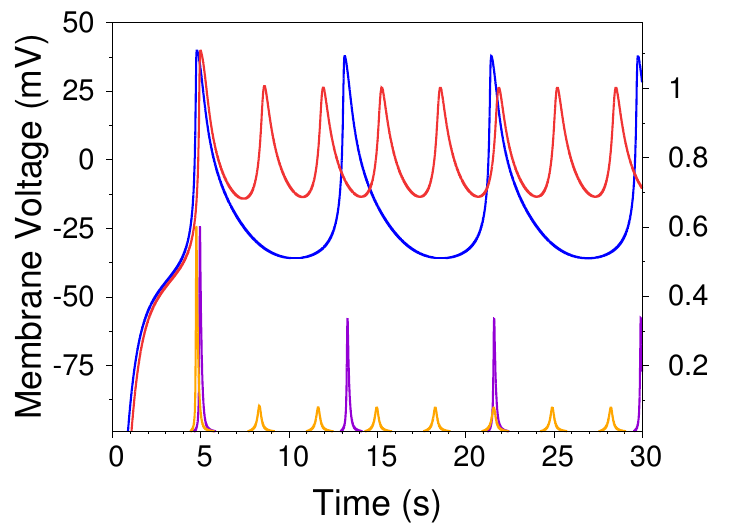}
	\caption{Two representative time traces of the voltage independent model, illustrating the sharp increase in frequency as one crosses from Region I to Region II in the phase diagram of Fig.~\ref{fig:pspace}. The blue trace has $k_{r} = 0.18 \, $s$^{-1}$ while the red trace has $k_{r} = 0.19 \, $s$^{-1}$, with all other parameters identical ($k_i = 10.4\,$s$^{-1}$). Firing is elicited by stepping $V_c$ from -200$\,$mV to -50$\,$mV at $t = 1\,$s.  The purple and orange traces show the probability that channels are open, $p_o$, for the blue and red traces respectively, with scale on the second y-axis. }
	\label{fig:time_trace_1}
\end{figure}

\begin{figure}
	\includegraphics[width=\linewidth]{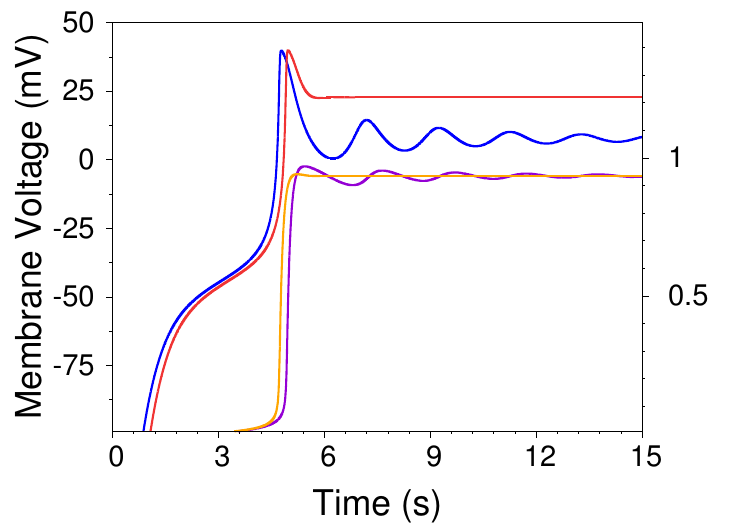}
	\caption{Two time traces of the voltage independent model, displaying damped oscillations and single shot AP. The blue trace ($k_{r} = 0.22\, $s$^{-1}$) corresponds to region III of the phase diagram of Fig.~\ref{fig:pspace}, the red trace  ($k_{r} = 0.5\, $s$^{-1}$) to region IV. The purple and orange traces display the corresponding probability that the channels are inactive, $p_i$. Parameters other than $k_r$ are identical to Fig.~\ref{fig:time_trace_1}.}
	\label{fig:time_trace_2}
\end{figure}

We identify four regions of distinct behavior. In region I the dynamical system (\ref{eq:membrane}), (\ref{eq:rates_1}) produces action potential trains (i.e. limit cycles). The firing rate increases for increasing $k_r$, while the width of the AP decreases for increasing $k_i$. Fig.~\ref{fig:time_trace_1} shows a corresponding time trace (blue). In region II the system exhibits ``oscillations'', distinct from Region I in that the amplitude is smaller, and the rate higher. The transition from region I to region II can be sharp, depending on the control parameters $\chi_c$ and $V_c$. The red trace in Fig.~\ref{fig:time_trace_1} displays this effect. We explore this transition in further detail in the next section. Region III corresponds to damped oscillations, the damping increasing for increasing $k_r$; a representative time trace is shown in Fig.~\ref{fig:time_trace_2} (blue). Finally, in region IV the system fires only once, after which the voltage remains constant at a relatively high value, corresponding to the limit cycle collapsing to a stable fixed point different from the resting potential. The red trace in Fig.~\ref{fig:time_trace_2} displays this behavior. \\

To summarize, for the case that the inactivation and recovery rates ($k_i$ and $k_r$) are voltage independent, we find four possible dynamic behaviors under constant current input: AP trains, oscillations, damped oscillations, and single shot firing. Remarkably, this phenomenology of the AA (one channel species with inactivation) is the same as for the Morris-Lecar system (two channel species without inactivation) \cite{Morris_Lecar,BifurcationsMorrisLecar2006,BifurcationAnalysisMorris2014}. \\ 

The phase diagram of Fig.~\ref{fig:pspace} is a slice through a higher dimensional parameter space with $V_c$ and $\chi_c$ held fixed. Changing these parameters (within a reasonable range) shifts the boundary lines in the $k_i$, $k_r$ plane but does not fundamentally alter the possible behaviors. Taking a different slice in parameter space (e.g. $\chi_c$ vs $V_c$) also yields the same regions of behavior.  \\

\begin{figure}
	\includegraphics[width=\linewidth]{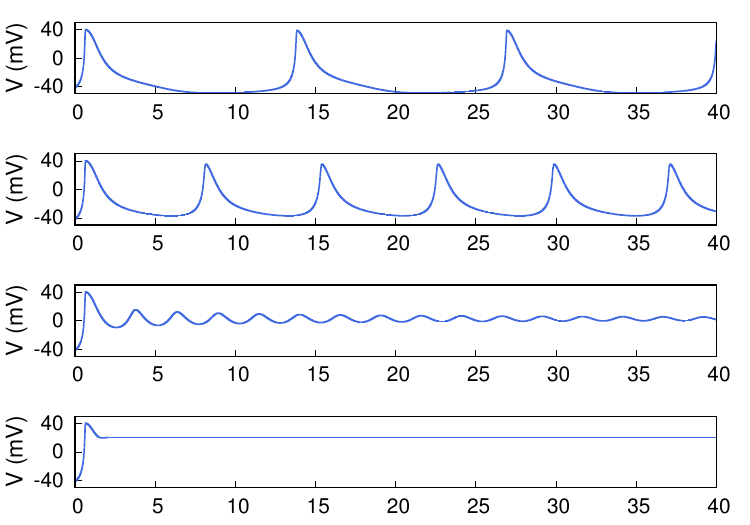}
	\caption{Representative time traces of the model with voltage dependent recovery and inactivation rates in the form of (\ref{eq:rates_2}). The x-axis is time (s), and the traces from top to bottom are representative of regions I$\,-\,$IV, and correspond to $\kappa_r = 0.037, 0.038, 0.06,$ and $0.3\,$s$^{-1}$, respectively. The fixed inactivation and recovery parameters are: $\kappa_i = 3\,$s$^{-1}$, $\alpha_r = -20\,$V$^{-1}$, $\alpha_i = 20\,$V$^{-1}$, $V_0^{(r)} = V_0^{(i)} = 0$, with clamp value $V_c = -56\,$mV. Other parameters ($N_0, C$, etc.) are identical to the voltage independent case (Fig.~\ref{fig:pspace}). The top two traces are chosen to show the sharp transition between Region I and Region II.} 
	\label{fig:vdep_vtraces}
\end{figure}

It is interesting to ask how the dynamics change if $k_i$ and $k_r$ are instead voltage dependent, as is the case for the KvAP channel used in the experiment. Taking the voltage dependence to be of the Arrhenius form, $k = \kappa \exp [\alpha (V-V_0)]$, simulations of the system show that the phenomenology remains the same.  The voltage dependent model produces the same four types of behavior (AP trains, oscillations, damped oscillations, and single shot AP) as we found in the voltage independent model, for a wide range of parameter values. Fig.~\ref{fig:vdep_vtraces} shows example time traces from a system with voltage dependent inactivation and recovery rates: $k_i = 3 \, e^{20V}$, $k_r = \kappa_r \, e^{-20V}$, $V_c = -56\,$mV, and all other parameters identical to the voltage independent model. In the figure, the system traverses through the four regions in the same fashion as before (I $\rightarrow$ II $\rightarrow$ III $\rightarrow$ IV) as $\kappa_r$ is increased; holding $\kappa_r$ fixed and varying another parameter again produces the same four regions of behavior.  Notably, the sharp transition between Region I and Region II which was found in the voltage independent model is preserved for certain parameter choices. \\

While the voltage dependent model more closely aligns with the situation of the experiment, it is not as useful for the purpose of analyzing the transitions between regions in the system, given that the voltage independent model has the same phenomenology with less parameters. Having shown that removing the voltage dependence from $k_i$ and $k_r$ does not fundamentally alter the available behaviors of the system, we proceed with further analysis using the voltage independent model.

\subsection{Bifurcations} 

\begin{figure} 
	\centering
	\subfigure[\label{fig:clvc_bif}]{\includegraphics[width=\linewidth]{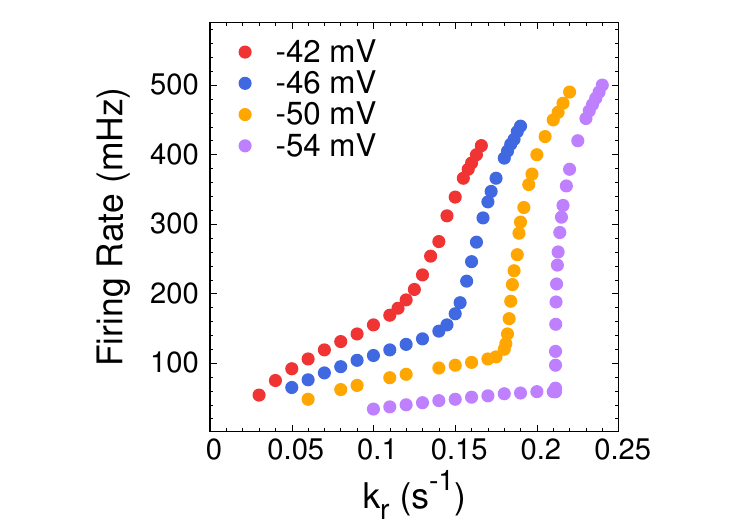}}
	\subfigure[\label{fig:k_oi_bif}]{\includegraphics[width=\linewidth]{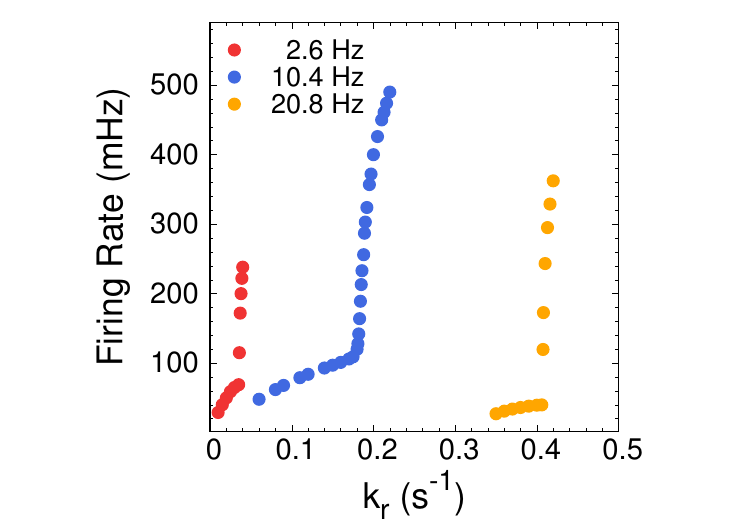}}  
	\caption{(a) Firing rate as a function of $k_r$ for several values of $V_c$ (legend) with $k_i = 10.4\,$s$^{-1}$. For each curve, the lower firing rate portion (to the left of the ``corner'') corresponds to Region I of the phase diagram, while the higher firing rate portion corresponds to Region II. In this example, the transition is sharp for $V_c = - 54 \, $mV. As $V_c$ is raised, the sharpness of the transition decreases and the transition region moves to smaller values of $k_r$. For each curve, the last 4 - 5 points at the higher frequencies lie in Region III (damped oscillations).  \\ 
	(b) Firing rate as a function of $k_r$, for different values of $k_i$ (legend) and fixed $V_c =  -50\,$mV. This plot displays the same transition as in (a). For both plots,  $\chi_c = 500\,$pS.}
	\label{fig:bif_1}
\end{figure}

\noindent We investigate the nature of the transition between regions I and II of the phase diagram of Fig.~\ref{fig:pspace}, for the voltage independent model. Fig.~\ref{fig:clvc_bif} shows the firing rate vs $k_r$ for various values of the clamp voltage $V_c$, with fixed $k_i = 10.4\,$s$^{-1}$, and $\chi_c = 500\,$pS. For $V_c = -54 \,$mV we see a sharp transition at $k_r \approx 0.21 \,$s$^{-1}$ where the firing rate increases steeply. This transition smoothes out as the clamp voltage is raised, with the transition point moving to lower values of $k_r$. Fig.~\ref{fig:k_oi_bif} displays the same transition for 3 different values of $k_i$, keeping $V_c = -50 \,$mV fixed. We see that the main effect of varying $k_i$ is to shift the transition point. \\ 

Displaying phase space trajectories often gives better insight into the nature of a bifurcation, however the dynamical system (\ref{eq:membrane}), (\ref{eq:rates_1}) is 3D, which makes phase space representations cumbersome.  We can make progress by noting that, in the regime $k_i , k_r \ll k_o , k_c$, the 3D system can be reduced to 2D without affecting the nature of the bifurcations \cite{Morris_Lecar}. We introduce a new coordinate $p_a (t)$, the probability that channels are ``active'' (i.e. not inactive: open or closed). Since the interconversion $C \rightleftharpoons O$ is ``fast'', we may make the substitution $p_o (t) \rightarrow p_a (t) P(V)$ in (\ref{eq:membrane}), where $P(V)$ is the equilibrium opening probability in the absence of inactivation, which is a function of voltage only. The function $P(V)$ is measured in the experiments \cite{Amila2013}; it is a Fermi-Dirac distribution (see (\ref{eq:rates_2})): 

\begin{equation}
P(V) = \frac{1}{k_c(V) / k_o (V) + 1} = \frac{1}{e^{- \frac{2 \alpha}{T}(V - V_0)} + 1}
\label{eq:open_prob}
\end{equation} 

\noindent Writing the total channel conductance $N_0 \chi$ as $\chi_0$ (so $\tau = C/\chi_0$), eq.~\ref{eq:membrane} can be made dimensionless by dividing all voltages by $V_N$ and scaling by $\tau$. With the condition $p_a (t) + p_i (t) = 1$, the system (\ref{eq:membrane}), (\ref{eq:rates_1}) can now be written in terms of the coordinates ($V(t)$, $p_a(t)$) as: 

%

\begin{equation}
	\begin{split}
		\frac{d V}{d t} = p_a(t) \, P[V(t)] \,  [1 - V(t)] + \chi_c \, \left [V_c - V(t) \right ] 
	\end{split}
	\label{eq:membrane_2D}
\end{equation}

\begin{equation}
	\begin{split}
		\frac{d p_a}{d t} = k_r - k_r \left[1 + \frac{k_i}{k_r} P[V(t)] \right] \, p_a(t)
	\end{split}
	\label{eq:rates_2D}
\end{equation}

\noindent where we have introduced dimensionless variables as mentioned before, dropping the tilde, so $V / V_N \rightarrow V$, $t /\tau \rightarrow t$, $\chi_c/\chi_0 \rightarrow \chi_c$, etc. Numerically, using ``standard values'' for our experimental parameters \cite{Ziqi2021}, $C = 300 \,$pF, $\chi = 200\,$pS, $N_0 = 100$, the time scale is $C / (N_0 \chi) = 1.5 \times 10^{-2} \,$s, so for example the (dimensional) rate $k_r = 0.2 \,$s$^{-1}$ corresponds to the dimensionless rate $(C k_r) / (N_0 \chi) = \tau k_r = 3 \times 10^{-3}$. If we regard the parameters ($\alpha$, $V_0$) which define the open probability function $P(V)$ as fixed, the 2D dynamical system (\ref{eq:membrane_2D}), (\ref{eq:rates_2D}) depends only on the four control parameters $\chi_c$, $V_c$, $k_r$, and $(k_i / k_r)$.

\begin{figure}
	\centering
	\subfigure[\label{fig:Limit_cycle_kr_6p0}]{\includegraphics[width=3.0in]{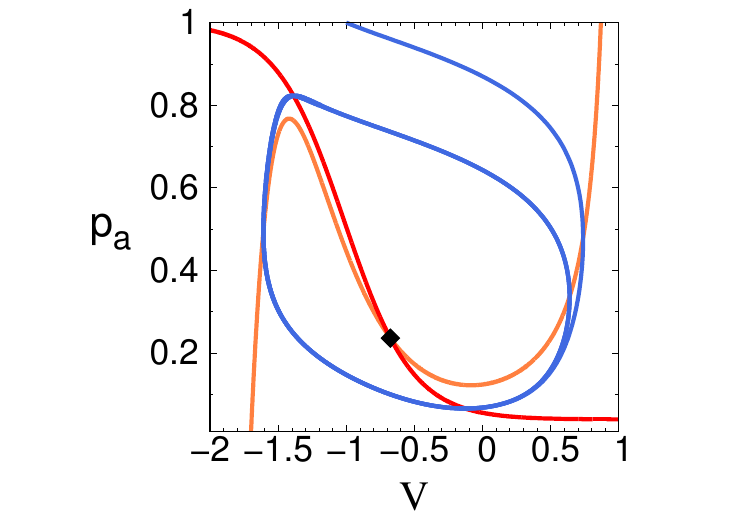}}
	\subfigure[\label{fig:AP_kr_6p0}]{\includegraphics[width=3.0in]{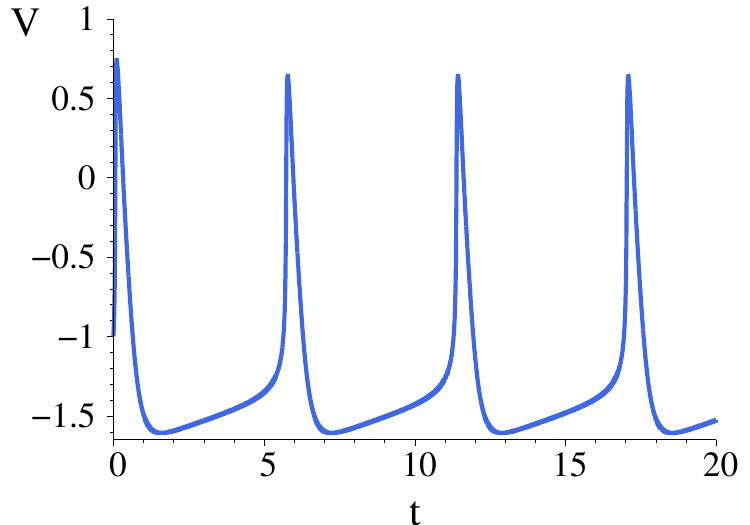}} 
	\caption{(a) Phase space trajectory (blue) in the $V$, $p_a$ plane for the 2D dynamical system (\ref{eq:membrane_2D}), (\ref{eq:rates_2D}),  displaying the limit cycle corresponding to an AP train, with $k_r = 6.0 \times 10^{-3}$, $k_i = 0.15$, $V_c = -1.7$, and $\chi_c = 0.05$, starting from the initial condition $(V = -1,p_a = 1)$. Also shown are the nullclines $dp_a / dt = 0$ (red) and $dV / dt = 0$ (orange); they intersect at an (unstable) fixed point.  (b) Time trace of the AP train corresponding to the limit cycle shown in (a); dimensionless $t$ and $V$ (see text).}
	\label{fig:phase_space_1}
\end{figure}

Referring to the phase diagram of Fig.~\ref{fig:pspace} as a guide, let us explore the nature of the transitions as we move along a horizontal line in that phase diagram (i.e. we vary $k_r$ at fixed $k_i$, other parameters fixed). For the fixed values $V_c = - 1.7$, $k_i = 0.15$, $\chi_c = 0.05$ we find the behavior of region I (AP trains) in the interval $4.03 \times 10^{-3} \lesssim k_r \lesssim 9.20 \times 10^{-3} $. Fig. \ref{fig:Limit_cycle_kr_6p0} shows a phase space trajectory for $k_r = 6.0 \times 10^{-3}$ (blue line), which is a limit cycle. The corresponding AP train is shown in Fig.~\ref{fig:AP_kr_6p0}. In Fig.~\ref{fig:Limit_cycle_kr_6p0} we also plot the nullclines $dp_a / dt = 0$ (red line) and $dV / dt = 0$ (orange line); they intersect at the unstable fixed point inside the limit cycle. \\

\begin{figure}
	\centering
	\subfigure[\label{fig:Spiral_kr_9p19}]{\includegraphics[width=3.0in]{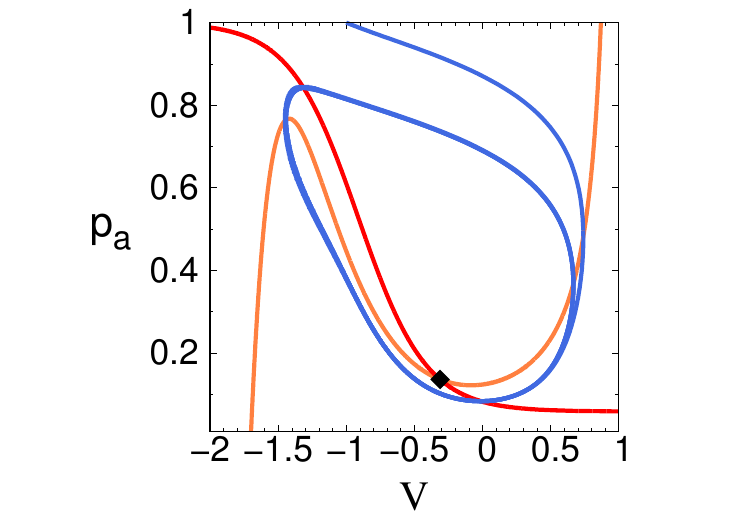}}  
	\subfigure[\label{fig:AP_kr_9p19}]{\includegraphics[width=3.0in]{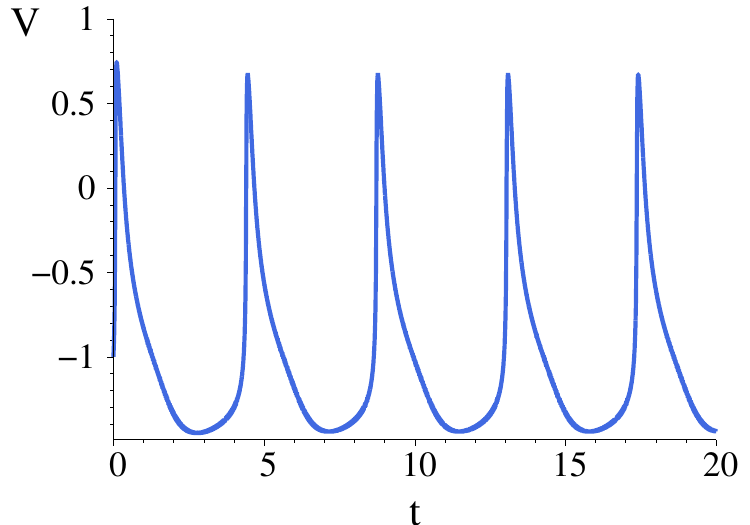}} 
	\caption{(a) Phase space trajectory displaying the limit cycle for $k_r = 9.20 \times 10^{-3}$  (other parameters are same as in Fig.~\ref{fig:phase_space_1}), just inside Region I.
	(b) Time trace of the AP train corresponding to the trajectory shown in (a).}
	\label{fig:phase_space_2}
\end{figure}

\begin{figure}
	\centering
	\subfigure[\label{fig:Spiral_kr_9p20}]{\includegraphics[width=3.0in]{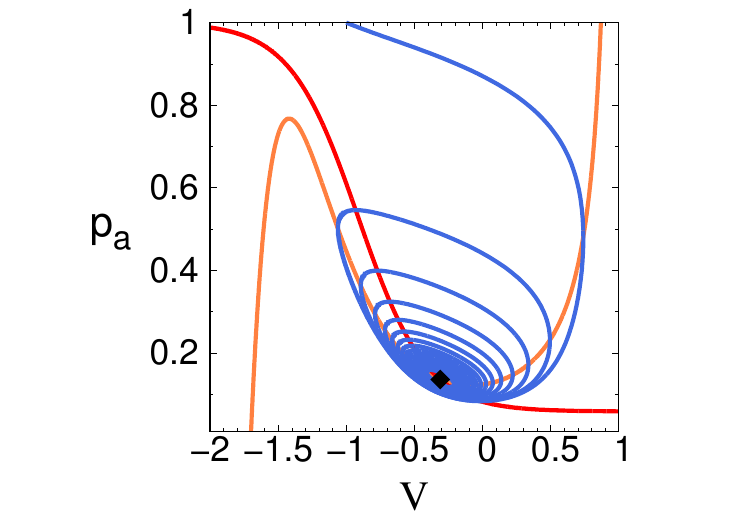}}
	\subfigure[\label{fig:Damped_oscill_kr_9p20}]{\includegraphics[width=3.0in]{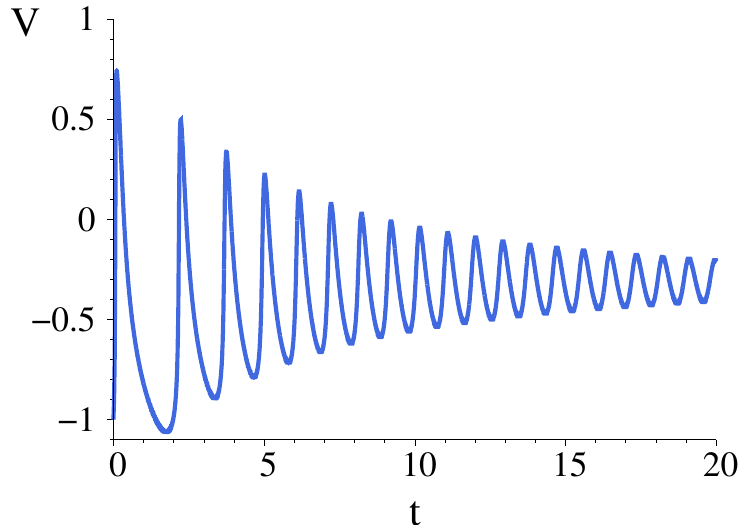}} 
	\caption{(a) Phase space trajectory for $k_r = 9.21 \times 10^{-3}$ (other parameters are same as in Fig.~\ref{fig:phase_space_1}), just outside Region I. There is no longer a stable limit cycle and the trajectory 
	spirals into the stable fixed point. (b) Time trace of the damped oscillations corresponding to the trajectory shown in (a). }
	\label{fig:phase_space_3}
\end{figure}

\begin{figure}
	\includegraphics[width=3in]{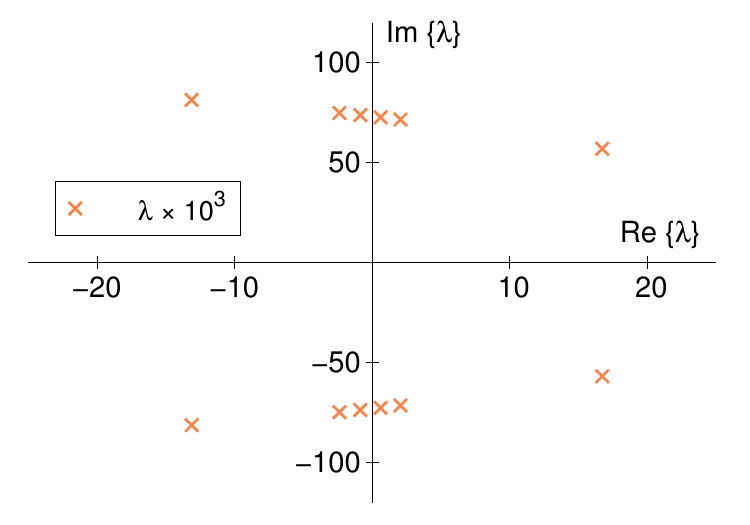}
	\caption{Eigenvalues $\lambda$ of the stability matrix at the fixed point, calculated numerically for the dynamical system (\ref{eq:membrane_2D}), (\ref{eq:rates_2D}), for different values of $k_r$. From left to right, the points correspond to: $k_r = (10.0, 9.30, 9.20, 9.10, 9.0, 8.0) \times 10^{-3}$. Other parameters are as in Fig.~\ref{fig:phase_space_1}. }
	\label{fig:eigenvalues}
\end{figure}

For $k_r \geq 9.21 \times 10^{-3}$ the system exhibits damped oscillations, corresponding to Region III in the phase diagram of Fig.~\ref{fig:pspace}. The transition between AP trains and damped oscillations is sharp. Fig.~\ref{fig:phase_space_2} shows the same quantities as Fig.~\ref{fig:phase_space_1}, for $k_r = 9.20 \times 10^{-3}$ (just inside the region of AP trains). Fig.~\ref{fig:phase_space_3} shows these plots for $k_r = 9.21 \times 10^{-3}$ (just outside the region of AP trains). Here the phase space trajectory spirals into the (stable) fixed point, corresponding to damped oscillations of $V(t)$. Note that with the above parameter values (specifically, the relatively small $k_i$) we did not cross Region II. Rather, we may say that the transitions $I \rightarrow II$ and $II \rightarrow III$ have ``merged'', in the sense that the transition from AP trains to damped oscillations is also accompanied by a steep increase in frequency (Figs.~\ref{fig:AP_kr_9p19}, \ref{fig:Damped_oscill_kr_9p20}). The phenomenology just described is that of a subcritical Hopf bifurcation \cite{Strogatz_Book}. The linear stability analysis of the fixed point close to the bifurcation shows that the eigenvalues of the stability matrix form a complex conjugate pair and cross the imaginary axis from right to left as the fixed point changes from unstable to stable. The eigenvalues $\lambda$ (real and imaginary part) are shown in Fig.~\ref{fig:eigenvalues} for different values of $k_r$ near the bifurcation. With the parameter values of this section, the bifurcation point is at $k_r \approx 9.21 \times 10^{-3}$. \\

For $k_r \gtrsim 12 \times 10^{-3}$ (not plotted) the system is so heavily damped that there are no oscillations; depending on initial conditions $V(t)$ either approaches the fixed point value from one side or fires once and then approaches the fixed point. This is the behavior of region IV of Fig.~\ref{fig:pspace}, the transition into this region being one from the underdamped to the overdamped regime. \\

\begin{figure}
	\centering
	\subfigure{\includegraphics[width=3.0in]{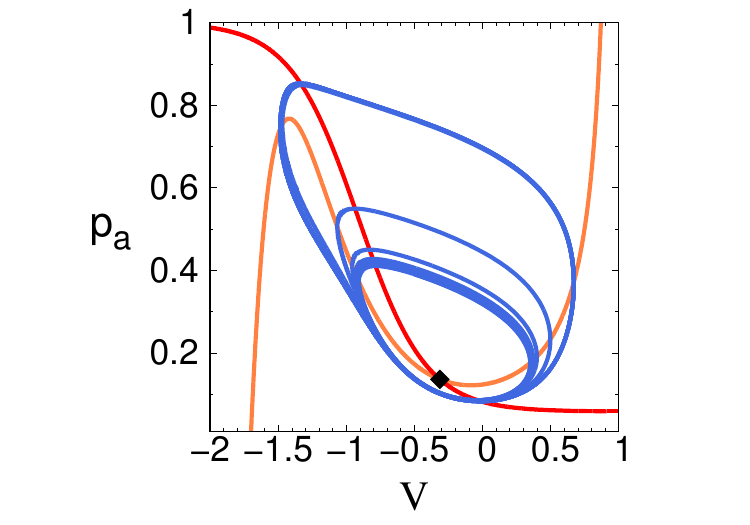}}
	\subfigure{\includegraphics[width=3.0in]{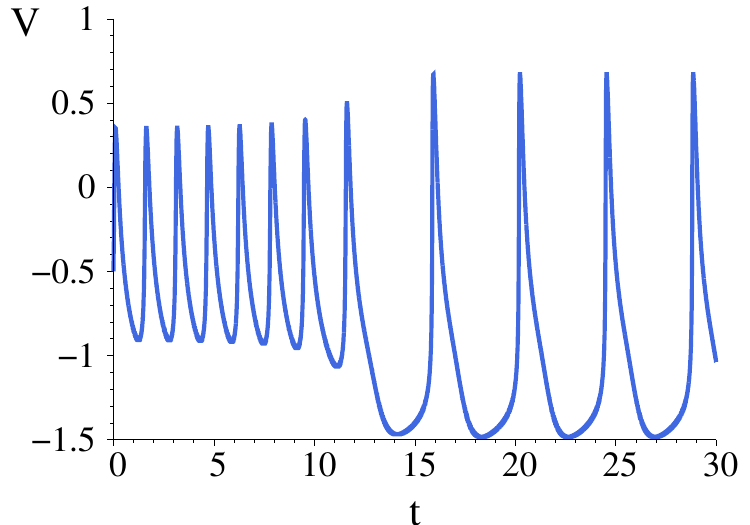}} 
	\caption{(a) Phase space trajectory for $k_r = 9.18 \times 10^{-3}$ and initial condition $(V = -0.5$, $p_a = 0.39)$, showing the presence of an unstable limit cycle inside the outer stable cycle. Other parameters are as in Fig.~\ref{fig:phase_space_1}. (b) Time trace of the trajectory shown in (a).}
	\label{fig:phase_space_4}
\end{figure}

In general, the behavior of the system is independent of the choice of initial conditions. Since there is only one fixed point, for any starting point $(V(0), p_a(0))$ the system either spirals into the fixed point, if it's stable, or moves to the limit cycle. However, there are exceptions to this near the Hopf bifurcation. Fig.~\ref{fig:phase_space_4} shows such a case. The system starts at $(V,p_a) = (-0.5, 0.39)$, which is close to the fixed point. The trajectory travels outwards and makes several loops before stabilizing at the larger stable limit cycle.  This behavior arises from the presence of an unstable limit cycle which is in between the stable fixed point and stable limit cycle. For $k_r$ values far from the Hopf bifurcation, this phenomenon is not seen because the fixed point itself is unstable. \\

\begin{figure}
	\centering
	\subfigure[\label{fig:pspace_2d_freq_vs_kr_for_Vc}]{\includegraphics[width=3.0in]{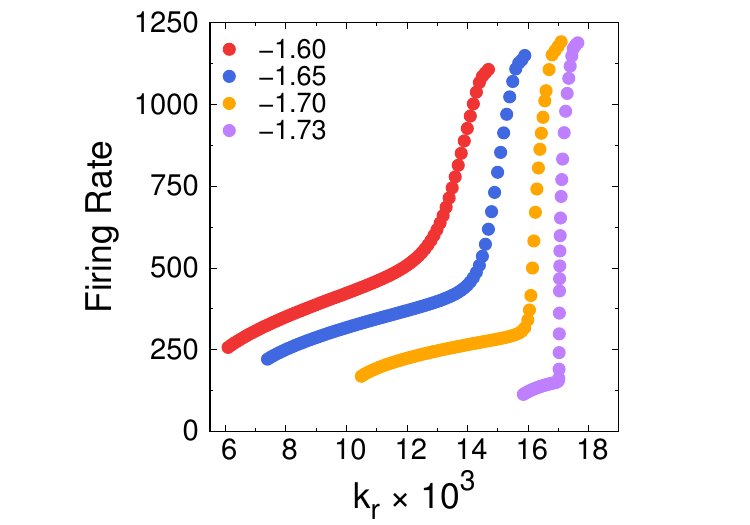}}
	\subfigure[\label{fig:pspace_2d_freq_vs_kr_for_ki}]{\includegraphics[width=3.0in]{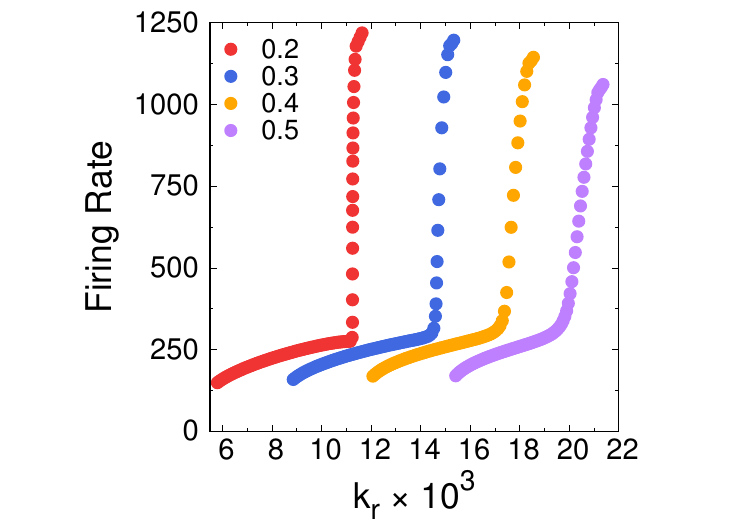}}
	\caption{(a) Firing rate for the reduced 2D model as a function of $k_r$, varying $V_c$ (legend) with $k_i = 0.35$
	and $\chi = 0.05$ held fixed. The phenomenology is the same as for the 3D system (Fig. \ref{fig:bif_1}), in particular, 
	there is a sharp transition for a critical value of $V_c$. (b) The same transition displayed for fixed $V_c = -1.7$, $\chi_c = 0.05$ and different values of $k_i$ (legend).}
	\label{fig:2d_transitions}
\end{figure}

Moving up in the phase diagram, i.e. increasing $k_i$, we recover Region II in the 2D system as well. 
This is displayed in Fig.~\ref{fig:2d_transitions}, which is analogous to Fig.~\ref{fig:bif_1}, but 
obtained for the 2D system. In Fig. \ref{fig:pspace_2d_freq_vs_kr_for_Vc} we plot the firing rate vs $k_r$ with $k_i$ fixed at $k_i = 0.35$ and different values of $V_c$. It is evident that exactly the same phenomenology occurs as for the 3D system: at a critical value of $V_c$ the transition is sharp ($V_c = - 1.73$ with these parameter values), and it smoothens out as $V_c$ is raised, with the transition shifting to smaller values of $k_r$. Fig. \ref{fig:pspace_2d_freq_vs_kr_for_ki} displays the same transition for $V_c = - 1.7$ fixed and different values of $k_i$ (legend). \\


\begin{figure}
	\centering
	\subfigure[\label{a}]{\includegraphics[width=3.0in]{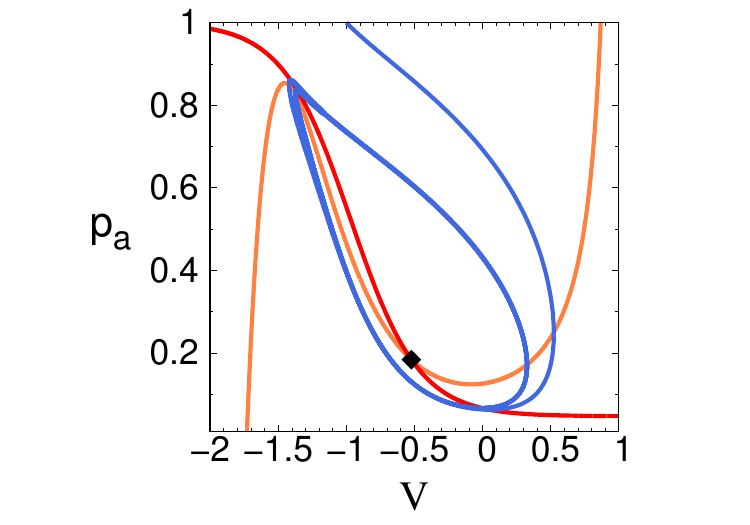}}
	\subfigure[\label{aa}]{\includegraphics[width=3.0in]{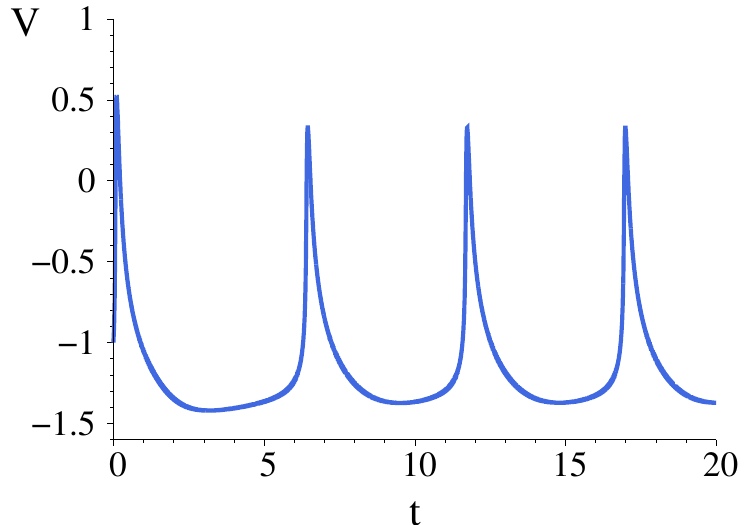}} 
	\caption{Phase space trajectory (a) and corresponding time trace (b) for the 2D system just prior to the I $\rightarrow$ II transition. $k_r = 17.03 \times 10^{-3}$, $k_i = 0.35$, $V_c = -1.73$, $\chi_c = 0.05$.}
	\label{fig:phase_trajectory_pre_transition}
\end{figure}

\begin{figure}
	\centering
	\subfigure[\label{aaa}]{\includegraphics[width=3.0in]{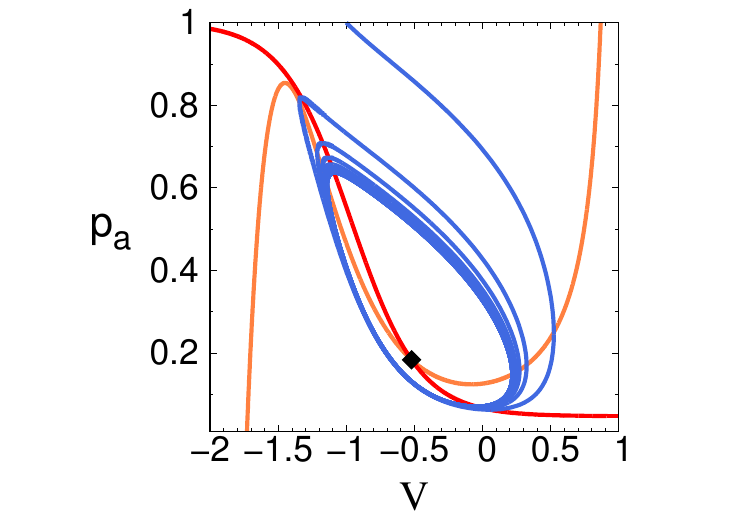}}
	\subfigure[\label{aaaa}]{\includegraphics[width=3.0in]{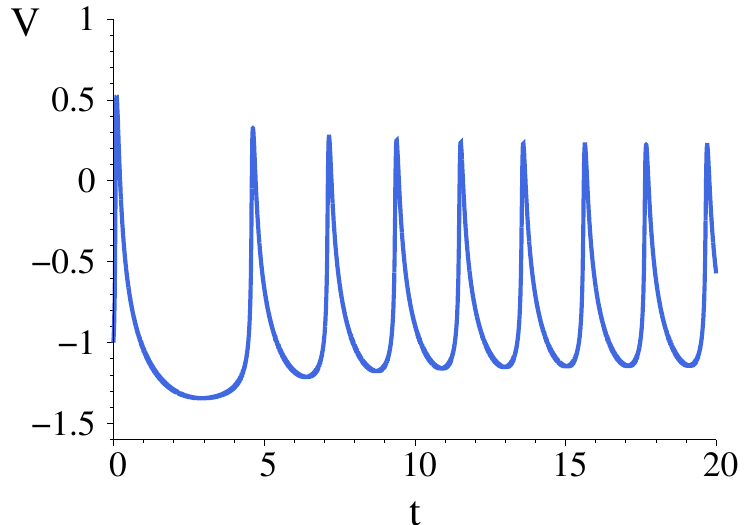}} 
	\caption{Phase space trajectory (a) and corresponding time trace (b) for the 2D system just after the I $\rightarrow$ II transition. $k_r = 17.05 \times 10^{-3}$, $k_i = 0.35$, $V_c = -1.73$, $\chi_c = 0.05$}
	\label{fig:phase_trajectory_post_transition}
\end{figure}

In Figs.~\ref{fig:phase_trajectory_pre_transition}, \ref{fig:phase_trajectory_post_transition} we show representative phase space trajectories and time traces across the Region I $\rightarrow$ Region II transition, for the 2D system. Different from the Hopf bifurcation corresponding to the transition II $\rightarrow$ III, the fixed point inside the limit cycle remains unstable on both sides of the transition. This is confirmed by an analysis of the eigenvalues of the stability matrix at the fixed point, which remain on the right side of the imaginary axis in both cases. 


In addition to the subcritical Hopf bifurcation, the system also contains a few other bifurcations which are similar to those of the Morris-Lecar system \cite{BifurcationsMorrisLecar2006, BifurcationAnalysisMorris2014}. Using $k_i$ as the bifurcation parameter, we briefly describe some of the bifurcations encountered as $k_i$ is increased. Starting at small $k_i$, there is one stable fixed point at the intersection of the $V$ and $p_a$ nullclines which is globally stable. As $k_i$ increases, the V nullcline moves to the right, and a saddle-node bifurcation will occur when the two nullclines intersect at a second point, which splits into two additional fixed points post-bifurcation. As $k_i$ is further increased, one of the newly created fixed points will annihilate with the original stable fixed point in another saddle-node bifurcation. This causes limit cycles to arise (Region I and Region II), as the only remaining fixed point is unstable. Finally, as $k_i$ is increased further, the remaining fixed point becomes stable through the Hopf bifurcation described above and all trajectories spiral into it (Region III), until eventually no oscillations occur (Region IV). \\

\begin{figure} 
	\centering
	\includegraphics[width=\linewidth]{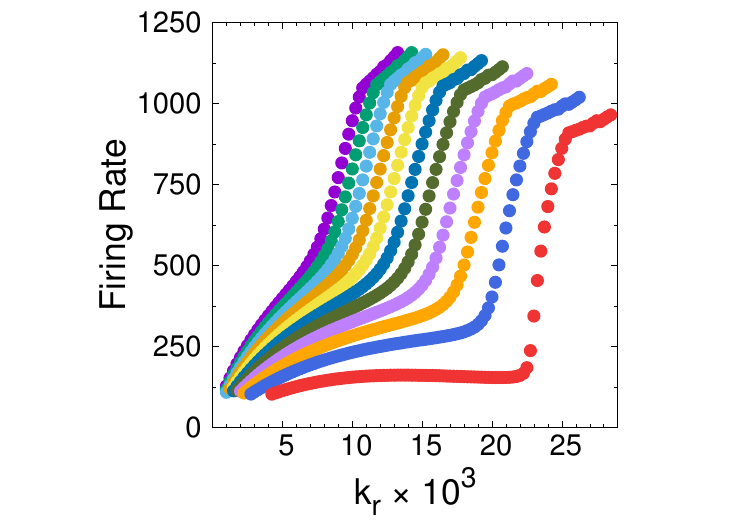}
	\caption{The transition from region I to region II in the 2D system, 
	explored along trajectories with fixed $k_i/k_r = 25$ (rather than fixed $k_i$ as in Fig.~\ref{fig:pspace_2d_freq_vs_kr_for_Vc}). The different curves correspond to $V_c$ in increments of 0.02, starting at $V_c = -1.52$ (violet) and ending at $V_c = -1.72$ (red).}
	\label{fig:pspace_2d_alt}
\end{figure}

The transitions described above may of course be explored following different trajectories in parameter space. Since for the 2D system (\ref{eq:membrane_2D}), (\ref{eq:rates_2D}) the control parameters which have to do with channel rates are $k_r$ and $k_i/k_r$, a natural trajectory is to keep the latter fixed. The overall picture remains the same: as an example, we show in Fig. \ref{fig:pspace_2d_alt} the firing rate vs $k_r$ for fixed $k_i/k_r$; the different curves correspond to the clamp values in the range $-1.72 < V_c < -1.52$, in increments of 0.02.


\subsection{Analogy to the Magnetization Transition} 

The plots of Fig.~\ref{fig:bif_1} and Fig.~\ref{fig:2d_transitions} present a qualitative resemblance to a number of equilibrium phase transitions. In Fig.~\ref{fig:clvc_bif}, the firing rate $\nu$ vs $k_r$ of the $V_c = - 54 \,$mV curve exhibits a sharp transition; for $k_r \ge k_r^c = 21.15 \times 10^{-2}\,$s$^{-1}$ we find power law behavior $(\nu - \nu_c) \propto (k_r - k_r^c)^{\beta}$ with $\nu_c \approx 58 \,$mHz and scaling exponent $\beta \approx 0.327$. For $V_c > - 54 \, $mV the transition appears smoothed out. This resembles the behavior of the magnetization $M$ vs temperature $T$ for a ferromagnet close to the Curie point. In zero external magnetic field ($H = 0$) the magnetization rises abruptly for $T < T_c$, exhibiting power law behavior $M \propto (T_c - T)^{\beta}$; experimentally, the scaling 
exponent for systems in the Ising universality class is $0.31 \le \beta \le 0.33$ ; for the Ising model in 3D it is 
$\beta \approx 0.325$ \cite{Nigel_Book}. \\
\noindent For finite field ($H \ne 0$) the transition appears smoothed out in the M - T plane. With the correspondence $\nu \leftrightarrow M$, $k_r \leftrightarrow T$, $V_c \leftrightarrow H$ the plots in Fig.~\ref{fig:bif_1} resemble the magnetization vs temperature as the external field is turned on. For the magnetic system, a plot $M$ vs $H$ would display the phenomenon of hysteresis for $T < T_c$. We therefore ask whether the firing rate $\nu$ vs clamp voltage $V_c$ 
could show hysteresis too, for certain values of $k_r$. 

\begin{figure}
	\centering
	\includegraphics[width=\linewidth]{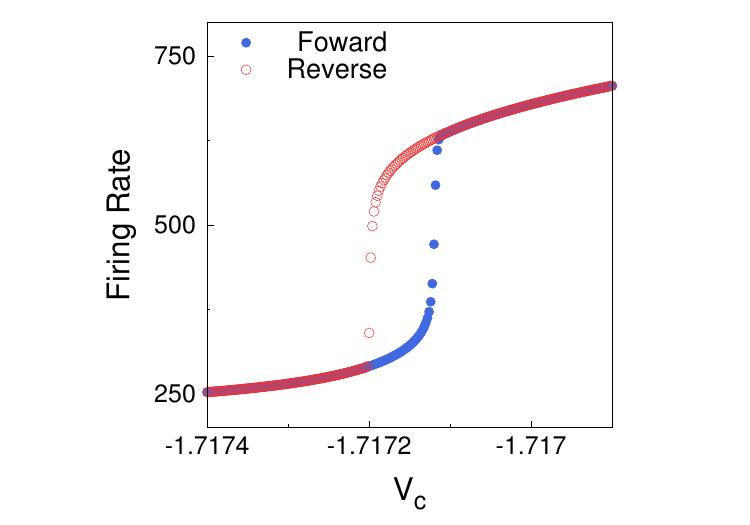}
	\caption{Hysteresis in the firing rate of the simulated 2D system (\ref{eq:membrane_2D}), (\ref{eq:rates_2D}).  Starting with $\chi_c = 0.05$, $k_r = 13.4 \times 10^{-3}$, $k_i = 0.25$, and $V_c^{(i)} = -1.718$, the clamp value is increased in increments of $2 \times 10^{-6}$ until $V_c^{(f)} = -1.716$ ($2,000$ total clamp values sampled), staying at each $V_c$ value for $t = 20$ so that a firing rate can be calculated. The process is then reversed, with the clamp returning to the initial value through the exact same intermediate values. The plot shows that the precise location 
of the transition between Region I and II depends on the direction of travel.}
	\label{fig:hysteresis}
\end{figure}

\noindent To investigate the occurrence of hysteresis in our model of the AA, we simulate the 2D voltage independent system in a slightly different way. We start the system in some initial state with $\chi_c$, $k_i$, and $k_r$ fixed, and an initial clamp value $V_{c}^{(i)}$. The clamp voltage is then increased (``adiabatically'') from this initial value  to a final value 
$V_{c}^{(f)}$ over a time interval $T$, in uniform increments ($V_{c}^{(f)} > V_{c}^{(i)}$).  The process is then reversed, with the clamp lowered from $V_{c}^{(f)}$ to $V_{c}^{(i)}$ over the time $T$.  The firing rate is calculated for each time increment $t$ ($t = T/N$ where $N$ is the number of $V_c$ values sampled between the initial and final values) and plotted as a function of $V_c$ for both the forward and reverse process. Using this protocol, we find that for certain parameter choices there is a difference in firing rate between the forward and reverse processes in the vicinity of the I $\rightarrow$ II transition, i.e. a hysteresis loop in the $V_c - \nu$ plane. Fig.~\ref{fig:hysteresis} shows the result for $\chi = 0.05$, $k_r = 13.4 \times 10^{-3}$, $k_i = 0.25$. The jump in firing rate corresponds to the I $\rightarrow$ II transition, and occurs at a slightly different $V_c$ depending on the direction the system approaches from. Note that the hysteresis loop shown here is not due to the subcritical Hopf bifurcation in the system, which corresponds to the transition II $\rightarrow$ III. The existence of hysteresis at a subcritical Hopf bifurcation is well known \cite{Strogatz_Book}; 
here it occurs at slightly larger values of $V_c$ (not shown on Fig.~\ref{fig:hysteresis}).

\section{Experimental Results}

\subsection{Measured Rates}
The purpose of mapping out the behavior of the AA dynamical system in parameter space is to provide guidance for the experiments in order to realize interesting dynamical behavior and understand the phenomenology. We are interested in knowing which of the dynamical phases can be accessed with the present experimental setup. Further, knowledge of the phase diagram in parameter space will guide the choice of alternative channels to improve the AA. To make progress, we need to know roughly where the present experimental system lies in the phase diagram described in section II. Though rates for the KvAP have been measured before \cite{MacKinnon_gatingModel2009}, these measurements are not easily mapped onto our system, as they are more complex (higher dimensional in parameter space) in order to account for the physical gating properties of the channels; whereas our model only has the minimal complexity needed 
to retain the core dynamics. \\

The equilibrium opening and closing rates of KvAP in the AA setting have been measured in a previous work \cite{Amila2013}, so we turn our attention to the inactivation and recovery rates $k_i$ and $k_r$. We measure these rates using a modified version of the system which is voltage clamped in the traditional manner (i.e. $R_{c}$ = 0).  The experiments are then carried out in standard electrophysiological fashion \cite{MacKinnon_gatingModel2009, hilleIonChannelsExcitable2001}, using voltage protocols adapted to obtain the ``effective'' rates of Fig.~\ref{fig:model}.  \\

\begin{figure}
		\subfigure[kdata][\label{fig:k_oi_data}]{\includegraphics[width=\linewidth]{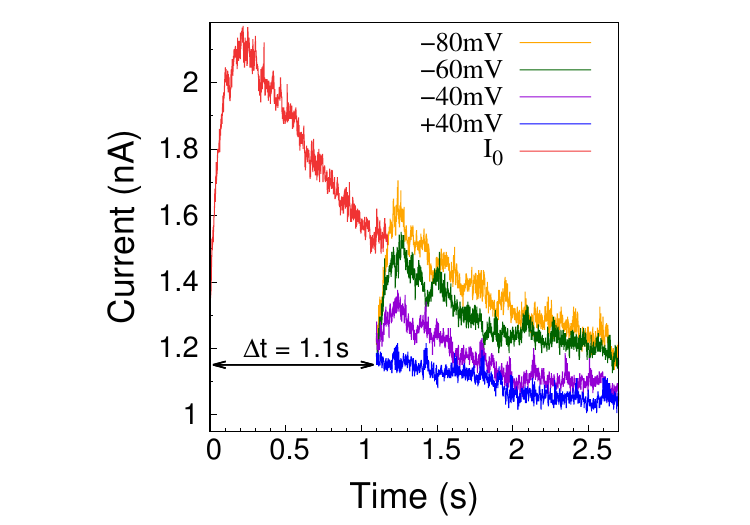}}
		\subfigure[kfit][\label{fig:k_oi_fit}]{\includegraphics[width=\linewidth]{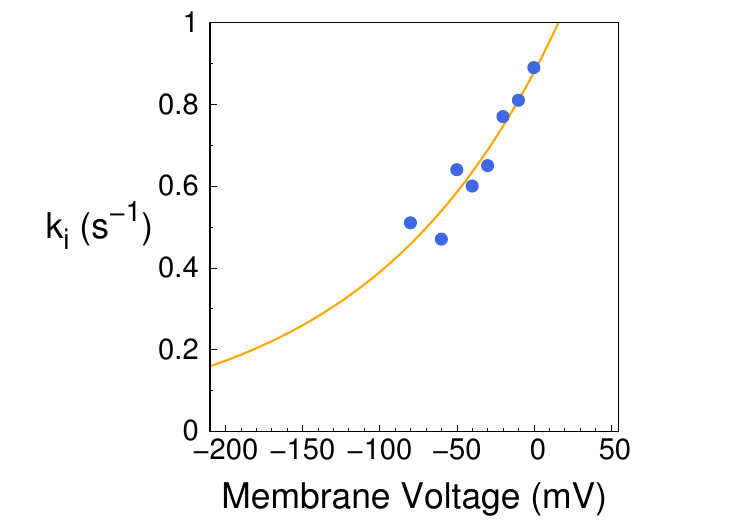}}
		\caption{(a) Representative time traces of the current corresponding to the voltage clamp protocol used to measure the inactivation rate $k_i (V)$. The initial peak (red trace) gives the maximum current $I_0$, and the ratio of subsequent peaks in comparison gives the ratio of inactivated channels after a time $t_2 = 1\,$s spent at the voltage indicated in the legend. ($\Delta t = t_1 + t_2$)\\  
		(b) The inactivation rate $k_i (V)$ plotted vs $V$, obtained from time traces as in (a). The solid line is a fit with an exponential function $k_i (V) = k_0 \, \exp(\beta V)$, returning the values $k_0 = 0.878\,$s$^{-1}$, $\beta = 8.13\,$V$^{-1}$.}
		\label{fig:k_oi} 
	\end{figure}

To measure $k_i (V)$, the system is first held at the resting voltage $V_r = -120 \,$mV, where almost all channels are in the closed state.  At $t = 0$ the voltage is stepped up to $V_1 = +100\,$mV and held there for a time $t_1 = 100 \,$ms. At the end of this time interval most channels are open and only few are inactivated, since at $V_1$ the opening rate is faster than $100 \,$ ms and the inactivation rate considerably slower. At $t = t_1$ the voltage is dropped to a lower (typically negative) value $V_2$, held there for a time $t_2$ ($ \sim 1 \, s$) before being stepped back up to $V_1 = +100$ mV. Finally the voltage is returned to the resting state $V_r$ in order to start another measurement. The measured quantity is the clamp current (equal to the channel current if we neglect leak currents). The proportion of open channels at time $t_1$ (immediately before the voltage is stepped to $V_2$) is constant. While the system is held at $V_2$, some channels will inactivate with a rate $k_i (V_2)$, thus the second step to $V_1$ will elicit a smaller current than the first.  The ratio of these two current peaks as a function of $V_2$ and $t_2$ allows us to extract the rate $k_i(V)$. Fig.~\ref{fig:k_oi_data} shows several current traces which illustrate the protocol. In formulas, the initial state is prepared with $p_i (t=0) \approx 0$ and $p_o (t=0) \approx 1$. Since we want the effective rate $O \rightarrow I$ we consider $(d/dt) p_o = - k_i (V) p_o$ for the dynamics while the system is held at $V = V_2$. Therefore after the time $t_2$ we have: $p_o (t_2, V) = p_o (0) \exp[- k_i (V) t_2]$. For a given voltage, the current is $I \propto p_o$ so $I_{peak} / I_0 = p_o (t_2, V) / p_o (0) = exp[- k_i (V) t_2]$ where $I_{peak}$ is the peak value of the current when the voltage is stepped to $V_1$ the second time, and $I_0$ the initial peak of the current, when the voltage is stepped to $V_1$ the first time (red trace in Fig.~\ref{fig:k_oi_data}). The quantity $-(1/t_2) \, \ln(I_{peak} / I_0) = k_i (V)$ (where $V=V_2$), obtained from traces as in Fig.~\ref{fig:k_oi_data}, is plotted vs $V$ in Fig. \ref{fig:k_oi_fit}, together with a fit to the form $k_i (V) = k_0 \, \exp(\beta V)$ (solid line), from which the parameters $k_0$ and $\beta$ are determined. \\

To measure the recovery rate $k_r$, the system is prepared in a state where all channels are inactive, which can be achieved by holding it at a high voltage $V_1 = 100 \,$mV.  The voltage is then stepped down to a $V_2$ below the threshold for firing (between $-120$ to $-80\,$mV), and held there for a time $\Delta t$; during this time, a fraction of the channels recover (into the closed state), with rate $k_r (V_2)$; then the voltage is stepped back up to $V_1$. A reference measurement is also taken where the system is held at $V_2$ for a long time ($> 20 \,$s) before stepping the voltage back to $V_1$. The ratio of measured current over reference current gives the ratio of open channels (i.e. channels which have recovered from inactivation) as a function of $V_2$ and $\Delta t$; $k_r (V)$ can be extracted by repeating the process for several values of $V_2$ and $\Delta t$. 
	
From these measurements, the resulting effective rates for the model of Fig. \ref{fig:model} are 
(in s$^{-1}$ ; $V$ in Volts):

$$k_i = 0.878 \times e^{8.13 \times V} \quad \quad k_r = 0.034 \times e^{-11.4 \times V}$$

These rates can be approximately mapped on to the voltage independent phase diagram of Fig.~\ref{fig:pspace} by considering the range of possible voltages of our system.  Typical experimental conditions are bounded from below by the resting voltage and above by the Nernst potential: -200$\,$mV $< V < $ 42$\,$mV. This corresponds to a window  $0.02 < k_r < 0.33$ and $0.17 < k_i < 1.24$ for the rates. This rectangle lies entirely in Region IV of the phase diagram of Fig.~\ref{fig:pspace}. This in agreement with the experimental observations, as no autonomous oscillations have been observed with the present system.  Thus, a single AA with the current setup based on the KvAP channel seems limited to single shot APs. In the next section we report some preliminary measurements on a system of two connected AAs, where we explore the feasibility of obtaining autonomous oscillations.   

\subsection{Connected AAs}

In the future it will be interesting to build networks of interconnected AAs. As a first step, we connected two AAs through electronic ``synapses''. Our synapse is a current clamp which takes as command voltage the voltage in the ``pre-synaptic'' axon and, if this voltage is above a set threshold, delivers a proportional current into the ``post-synaptic'' axon (similar to a much simplified version of the ``dynamic clamp'' used in some electrophysiology experiments \cite{DynamicClampPowerful2006}). Thus a synapse is characterized by two parameters: the threshold voltage $V_T$, and the coefficient of proportionality $\alpha$ between input voltage and output current. A synapse connecting AA1 and AA2 delivers a current into AA2 given by $I_2(t) = \alpha \, V_1(t) \, \Theta [V_1(t) - V_T]$ where $\Theta$ is the step function. For $\alpha > 0$ the synapse is ``excitatory'' while for $\alpha < 0$ it is inhibitory. In the following experiments we keep the threshold at $V_T = 0$, while typical values of the synapse ``strength'' are $|\alpha| \sim 10 \,$pA / mV $= 10\,$nS. \\

\begin{figure}
	\includegraphics[width=\linewidth]{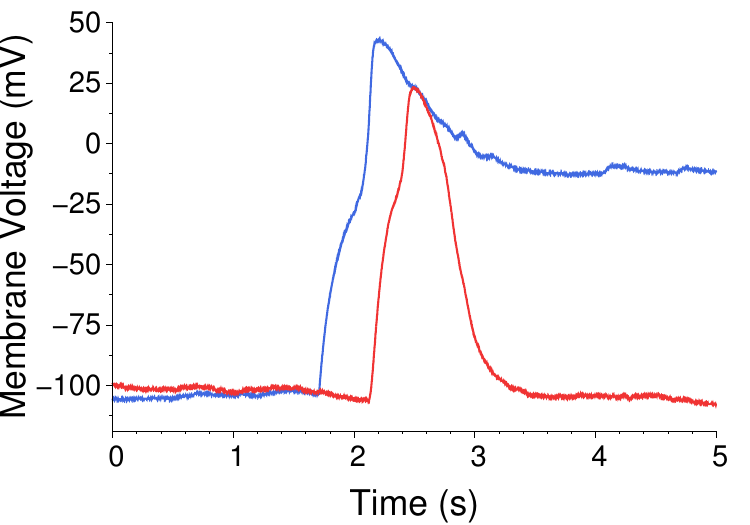}
	\caption{Time traces from two AAs connected by one excitatory synapse. Blue is $V_1$ (the voltage in AA1), 
	red is $V_2$. AA1 is caused to fire by raising its CLVC above threshold (to -20 mV); the CLVC of AA2 is held constant at -100 mV. As $V_1$ crosses zero, the synapse starts to inject current into AA2, eventually causing it to fire. As $V_1$ re-crosses zero downwards, the synapse shuts off and $V_2$ is pulled back to the initial resting potential by its CLVC and inactivation. The end result is a complete action potential cycle for AA2, which returns to its initial ``resting'' state.  }
	\label{fig:pushhold}
\end{figure}

The simplest system that can be made with this setup consists of two AAs connected by a single ``excitatory'' synapse ($\alpha > 0$). Fig.~\ref{fig:pushhold} shows experimental time traces of the voltage in the two axons for this configuration. The first axon (AA1: blue line, $V_1$) is caused to fire in the standard way, by stepping up its clamp (CLVC1) from the resting value to a value above threshold; as $V_1$ crosses zero, the synapse starts to inject positive current into AA2, eventually causing it to fire (AA2: red line, $V_2$). As channels inactivate, $V_1$ crosses zero again in the downwards direction, the synapse stops injecting current, and $V_2$ is pulled back to the resting potential by a combination of inactivation and its clamp (CLVC2). During this whole process CLVC2 is held steady at the resting value. $V_1$ does not come back to the resting potential because there are no further inputs to CLVC1 after the initial step up. The end result is that AA2 goes through a complete action potential cycle, including repolarization, with AA1 acting as the input. If AA2 was connected in the same way to a third axon AA3, a similar action potential cycle would be generated in AA3, and so on. A system of several AAs linked in such a way would allow for discrete spatial propagation of action potentials. This configuration is similar to a previously reported result \cite{Hector2017} in which the firing of an AP in AA1 caused firing in AA2, the only difference being that AA1 is made to fire via adjustment of its CLVC rather than using an external current source. To summarize: in this configuration, AA1 provides an input signal to AA2, which then fires a complete action potential cycle. We could think of AA1 as a sensory input (which could be realized in practice by embedding light or chemically gated channels in AA1, for example). \\ 
 
\begin{figure}
	\centering
	\subfigure[Push_Pull_exp][\label{fig:Push_Pull_exp}]{\includegraphics[width=3in]{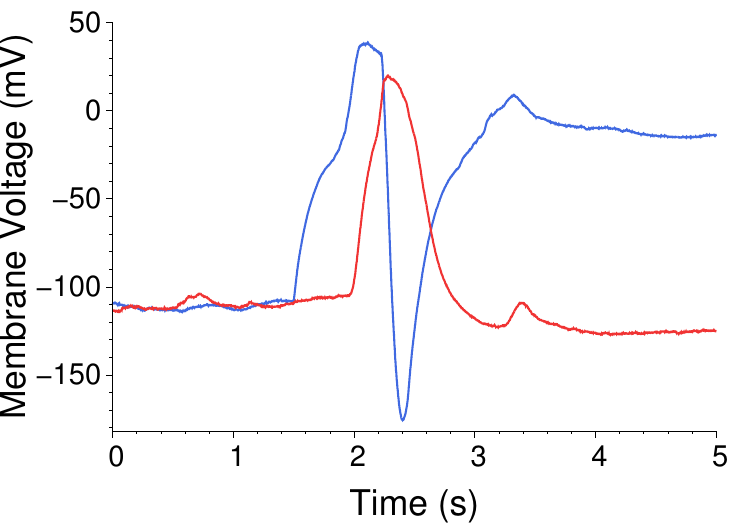}}
	\subfigure[Push_Pull_sim][\label{fig:Push_Pull_sim_V}]{\includegraphics[width=3in]{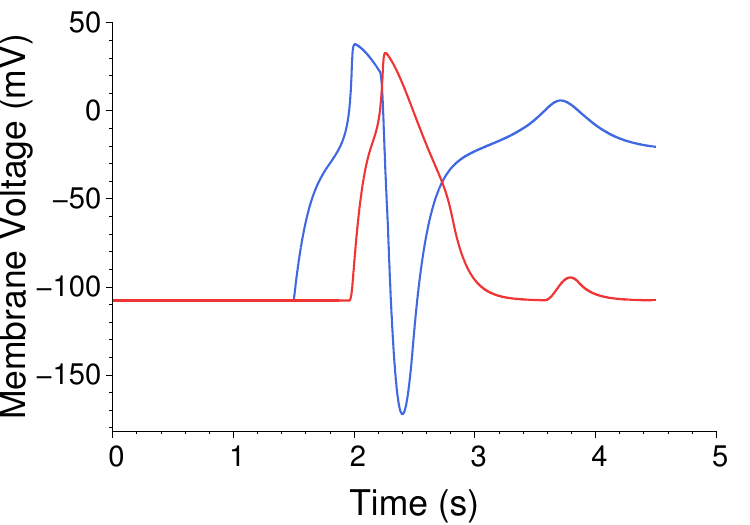}}
	\caption{(a) Time traces of the voltage from two AAs connected by one excitatory and one inhibitory synapse, in a feedback loop. To initiate the process, CLVC1 is raised above threshold at $t \approx 1.5 \,$s and then held constant; 
CLVC2 is held fixed at $V_r \approx -100\,$mV throughout. Firing of AA1 (blue trace) causes a positive synaptic current ($\alpha_{12} = 6\,$nS, $V_T = 0$) which induces firing of AA2 (red trace). When AA2 fires, the negative synaptic current ($\alpha_{21} = -25\,$nS, $V_T = 0$) injected in AA1 pulls $V_1$ down sharply. In this experiment, not enough channels in AA1 recovered from inactivation (during the negative voltage swing) in order for AA1 to fire again, which would lead to sustained oscillations. \\ 	
	(b) Numerical simulation of the system in (a), using the (voltage dependent) model (\ref{eq:membrane}), (\ref{eq:rates_1}), (\ref{eq:rates_2})  with a combination of measured rates and fitted parameters: $N_0 = 250$, $C = 275\,$pF, $\chi = 167\,$pS, $\alpha_{12} = 7.33\,$nS, $\alpha_{21} = -10.67\,$nS, and $V_T = 0$. Note that the discrepancy in synapse strengths is due to a large leak current $\chi_\ell$ in the experiment.}
	\label{fig:pushpull} 
\end{figure}

As shown in the previous section, as a result of the inactivation dynamics of the KvAP channel, a single AA does not sustain autonomous oscillations for a constant input current. Nevertheless, with two such AAs it is in principle possible to construct an oscillator. For this purpose we add a second synapse to the previous construction, providing inhibitory feedback. Now AA1 connects to AA2 through an excitatory synapse ($\alpha > 0$) and AA2 connects back to AA1 through an inhibitory synapse ($\alpha < 0$). Fig.~\ref{fig:Push_Pull_exp} shows corresponding experimental time traces (blue trace: $V_1$; red trace: $V_2$).  \\

\begin{figure}
	\centering
	\includegraphics[width=\linewidth]{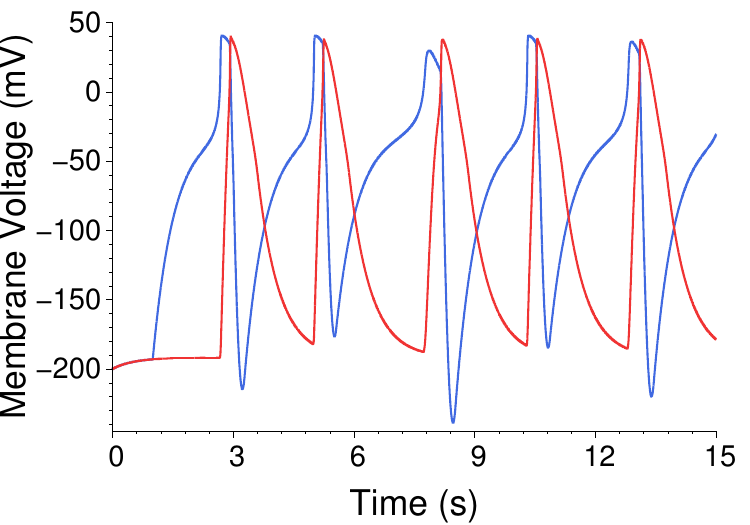}
	\caption{Time trace from a simulation of the two axon system, showing autonomous oscillations (AP trains). Blue trace is $V_1$, red trace $V_2$. Parameters are the same as for Fig.~\ref{fig:Push_Pull_sim_V}, except the inhibitory synapse strength has been increased to $\alpha_{21} = -20\,$nS. The resulting increased hyperpolarization of AA1 allows more channels to recover from inactivation, resulting in sustained oscillations. }
	\label{fig:sim_osc} 
\end{figure}

The rising part is similar to Fig.~\ref{fig:pushhold}, however now when $V_2$ crosses zero, the inhibitory synapse starts injecting negative current into AA1, pulling $V_1$ down to negative values below the resting potential (``hyperpolarization''). As $V_2$ crosses zero again on the falling edge, the synapse shuts off and AA1 repolarizes ($V_1$ rises again) since the clamp CLVC1 is kept steady at the stepped up value (i.e. constant stimulus conditions for AA1). Now in principle the process could repeat and generate a train of APs, i.e. an oscillator. Specifically, if the hyperpolarizing step lasts a sufficient amount of time, depending on synapse strength and channel inactivation/recovery rates, then when the action potential in axon 2 ceases, the voltage in axon 1 will return to close to the CLVC1 value (which is above threshold) and fire again. This does not quite happen in the experiment shown (but notice the little blip in the traces at $t \approx 3.3 \,$s, which is a partial firing of the system), for the reason that too many channels in AA1 are still inactivated to fire a full AP. In Fig.~\ref{fig:Push_Pull_sim_V} we show time traces for $V_1$ and $V_2$ from a numerical simulation of the system, using the voltage dependent model (\ref{eq:membrane}), (\ref{eq:rates_1}), (\ref{eq:rates_2}) for the individual axons. We used the experimentally measured rates (see previous section) and fitted the remaining parameters ($N_0, C, \alpha_1, \alpha_2$). It is apparent that the model reproduces the dynamics of the real system quite well. We may then interrogate the model on the conditions for obtaining oscillations. It turns out that adjusting synapse strength is enough. Fig.~\ref{fig:sim_osc} shows autonomous oscillations in the model, obtained with the same parameter settings as for Fig.~\ref{fig:Push_Pull_sim_V}, except the strength of the inhibitory synapse AA2 $\rightarrow$ AA1 has been increased from $\alpha_{21} = -10.67\,$nS to $\alpha_{21} = -20\,$nS. We conclude that autonomous oscillations are achievable with the current experimental system (though we have not yet been able to obtain them). The key factor in determining whether oscillations are sustained or die out (as in the experiment of Fig.~ \ref{fig:Push_Pull_exp}) is how low AA1 is pulled by the negative feed back from AA2. A hyperpolarization value for AA1 of $\lesssim -200$\,mV is indicated by the simulations to be the minimum required for sustained oscillations under the present conditions.

\section{Discussion}

The simplified Hodgkin-Huxley type model we use captures the dynamics of the experimental system quite well, 
with a minimum number of parameters. This is shown in Section III B where we compare model and experiments for an elaborate experimental system consisting of two AAs connected by ``synapses''. The reduced number of parameters in the model allows us to map out important features of the system's phase diagram in parameter space (Fig. \ref{fig:pspace}). The main conclusion is that the AA, a synthetic biology system consisting of one voltage gated channel species with inactivation, is dynamically equivalent to  the biological system of two voltage gated channel species without inactivation (the Morris-Lecar system). Everything the Morris-Lecar system can do, the AA can do, in principle. This raises the question of whether action potentials dependent on a single gated channel species exist (or have existed) in nature. As far as we know, no such system has been identified thus far.\\ 

We have discussed in detail some of the bifurcations which take the system from one region of the phase diagram to another; these have a universal character, which therefore should be maintained across different systems, from the AA to the Morris-Lecar dynamical system to the barnacle muscle fiber to the rat neuron. Indeed, the Hopf bifurcation corresponding to the onset of AP trains, which here we discuss for the AA (Section II C), is well established for the neuron \cite{Koch_Book}. Less established is the transition separating regions I and II in the phase diagram, in fact we do not find that it is discussed in the literature either in theory or experiments. A qualitative analogy with the magnetization transition also prompted us to look for hysteresis, which is indeed present. \\

\begin{figure}
	\includegraphics[width=\linewidth]{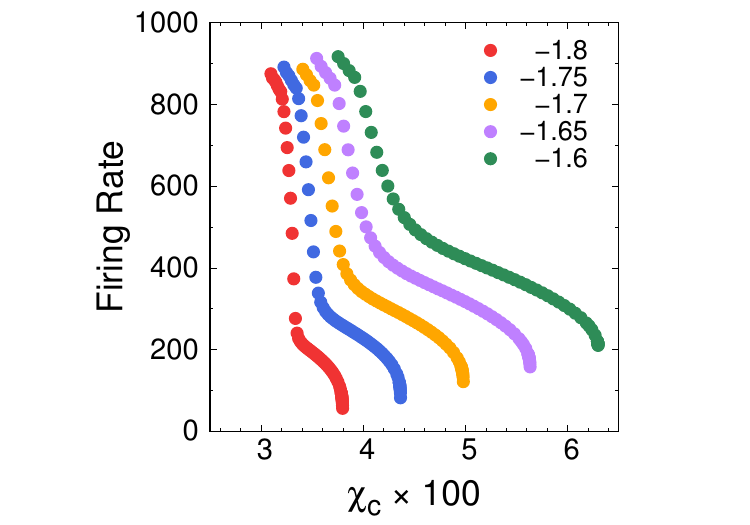}
	\caption{Firing rate of the AA as a function of the clamp conductance, $\chi_c$, computed from the 2D voltage independent model.  Each curves corresponds to a different $V_c$ values (legend). For lower clamp values a sharp transition occurs in firing rate as $\chi_c$ is increased past a critical value (Region II $\rightarrow$ Region I). Past the transition, the firing rate continues to slowly decrease until a discontinuity occurs due to the clamp current overwhelming the channel current (no firing).}
	\label{fig:chi_vs_freq}
\end{figure}

We have discussed these bifurcations mainly as a function of the inactivation and recovery rates of the channels, $k_i$ and $k_r$, corresponding to the phase diagram of Fig.~\ref{fig:pspace}. The reason is to establish guidelines for the future choice of different channels to improve the experimental system. The main conclusion here is that we want a channel with much faster (or more strongly voltage dependent) inactivation. At the same time, it should be noted that the same bifurcations can instead be explored as a function of parameters which are experimentally controlled in the AA. For example, Fig.~\ref{fig:chi_vs_freq} shows plots of the firing rate vs clamp conductance $\chi_c$ for the 2D model of Section II C, obtained from the simulation for different values of the clamp voltage $V_c$. The transition Region I $\rightarrow$ Region II is again visible as a sharp increase in firing rate as $\chi_c$ is lowered past a critical value. This is the same transition as in Figs.~\ref{fig:bif_1} and \ref{fig:2d_transitions}. The fact that this transition is also present in this slice of the phase diagram means that it may be possible to design an experiment to observe this transition in the AA, as both $\chi_c$ and $V_c$ are control parameters.  It would be difficult to follow this behavior in experiments with living cells, because one does not have the same control over the experimental parameters. Also, a bifurcation which is sharp in the model may appear smoothed out in the more complex environment of the cell. However, the dynamics of the biological system is likely to retain a signature of the underlying bifurcation, which therefore may provide a way to classify the behavior. \\ 

Finally, we present experiments where we connect two AAs through electronic ``synapses'' - a step towards constructing networks. In the configuration corresponding to Fig.~\ref{fig:pushhold}, a step perturbation of AA1 (the ``input'') evokes a single shot AP in AA2 (the ``output''). In terms of signal processing, suppose AA1 was a light sensitive channel, and the input consisted of a (slowly) blinking light. The system would encode each event of the light turning on into a single standardized AP (independent of the duration of the ``on'' phase), which could be further propagated down a network. By connecting two AAs in a feedback loop, it should be possible to construct an autonomous limit cycle oscillator. Fig.~\ref{fig:Push_Pull_exp} displays an attempt which was not quite successful, but indicates the feasibility of such a system. The difficulty lies in the fragility of the experimental system, which makes it difficult to tune parameters ``on the fly''. Our plan going forward is to both demonstrate this oscillator and also improve robustness. \\ 

The parameters which are currently directly controlled in the experiments are the clamp voltage and clamp conductance. Though much more cumbersome, the inactivation and recovery rates for a given channel could in principle be modified as well.  Previous work has shown that different compositions of the lipid membrane has an affect on the kinetics of the channel \cite{MacKinnon_gatingModel2009}, the caveat being that there is no quantitive way of knowing how the rates will change in response to a change in membrane composition.  Channel kinetics also change as a function of temperature \cite{ranjanKineticMapHomomeric2019}, which could be another way to indirectly tune the system.  A combination of such methods may be an effective strategy for exploring the parameter space experimentally. To expand the system beyond a two axon setup, further work is needed due to the difficulty in maintaining multiple functioning AAs, owing to the fragility of the lipid membrane setup.  Possible strategies include using hydrogels to stabilize the system \cite{jeonHydrogelEncapsulatedLipidMembranes2006, jeonBlackLipidMembranes2008}, or perhaps moving from a suspended lipid system to a supported lipid system.


\begin{acknowledgments}

\end{acknowledgments}

\bibliography{Ziqi_2}

\begin{thebibliography}{21}%
\makeatletter
\providecommand \@ifxundefined [1]{%
 \@ifx{#1\undefined}
}%
\providecommand \@ifnum [1]{%
 \ifnum #1\expandafter \@firstoftwo
 \else \expandafter \@secondoftwo
 \fi
}%
\providecommand \@ifx [1]{%
 \ifx #1\expandafter \@firstoftwo
 \else \expandafter \@secondoftwo
 \fi
}%
\providecommand \natexlab [1]{#1}%
\providecommand \enquote  [1]{``#1''}%
\providecommand \bibnamefont  [1]{#1}%
\providecommand \bibfnamefont [1]{#1}%
\providecommand \citenamefont [1]{#1}%
\providecommand \href@noop [0]{\@secondoftwo}%
\providecommand \href [0]{\begingroup \@sanitize@url \@href}%
\providecommand \@href[1]{\@@startlink{#1}\@@href}%
\providecommand \@@href[1]{\endgroup#1\@@endlink}%
\providecommand \@sanitize@url [0]{\catcode `\\12\catcode `\$12\catcode
  `\&12\catcode `\#12\catcode `\^12\catcode `\_12\catcode `\%12\relax}%
\providecommand \@@startlink[1]{}%
\providecommand \@@endlink[0]{}%
\providecommand \url  [0]{\begingroup\@sanitize@url \@url }%
\providecommand \@url [1]{\endgroup\@href {#1}{\urlprefix }}%
\providecommand \urlprefix  [0]{URL }%
\providecommand \Eprint [0]{\href }%
\providecommand \doibase [0]{https://doi.org/}%
\providecommand \selectlanguage [0]{\@gobble}%
\providecommand \bibinfo  [0]{\@secondoftwo}%
\providecommand \bibfield  [0]{\@secondoftwo}%
\providecommand \translation [1]{[#1]}%
\providecommand \BibitemOpen [0]{}%
\providecommand \bibitemStop [0]{}%
\providecommand \bibitemNoStop [0]{.\EOS\space}%
\providecommand \EOS [0]{\spacefactor3000\relax}%
\providecommand \BibitemShut  [1]{\csname bibitem#1\endcsname}%
\let\auto@bib@innerbib\@empty
\bibitem [{\citenamefont {Gerstner}\ \emph {et~al.}(1997)\citenamefont
  {Gerstner}, \citenamefont {Kreiter}, \citenamefont {Markram},\ and\
  \citenamefont {Herz}}]{gerstner1997}%
  \BibitemOpen
  \bibfield  {author} {\bibinfo {author} {\bibfnamefont {W.}~\bibnamefont
  {Gerstner}}, \bibinfo {author} {\bibfnamefont {A.~K.}\ \bibnamefont
  {Kreiter}}, \bibinfo {author} {\bibfnamefont {H.}~\bibnamefont {Markram}},\
  and\ \bibinfo {author} {\bibfnamefont {A.~V.~M.}\ \bibnamefont {Herz}},\
  }\href {https://doi.org/10.1073/pnas.94.24.12740} {\bibfield  {journal}
  {\bibinfo  {journal} {Proceedings of the National Academy of Sciences}\
  }\textbf {\bibinfo {volume} {94}},\ \bibinfo {pages} {12740} (\bibinfo {year}
  {1997})}\BibitemShut {NoStop}%
\bibitem [{\citenamefont {Ariyaratne}\ and\ \citenamefont
  {Zocchi}(2016)}]{Amila2016}%
  \BibitemOpen
  \bibfield  {author} {\bibinfo {author} {\bibfnamefont {A.}~\bibnamefont
  {Ariyaratne}}\ and\ \bibinfo {author} {\bibfnamefont {G.}~\bibnamefont
  {Zocchi}},\ }\href@noop {} {\bibfield  {journal} {\bibinfo  {journal} {J.
  Phys. Chem. B}\ }\textbf {\bibinfo {volume} {120}},\ \bibinfo {pages} {6255}
  (\bibinfo {year} {2016})}\BibitemShut {NoStop}%
\bibitem [{\citenamefont {Vasquez}\ and\ \citenamefont
  {Zocchi}(2017)}]{Hector2017}%
  \BibitemOpen
  \bibfield  {author} {\bibinfo {author} {\bibfnamefont {H.~G.}\ \bibnamefont
  {Vasquez}}\ and\ \bibinfo {author} {\bibfnamefont {G.}~\bibnamefont
  {Zocchi}},\ }\href@noop {} {\bibfield  {journal} {\bibinfo  {journal} {EPL}\
  }\textbf {\bibinfo {volume} {119}},\ \bibinfo {pages} {48003} (\bibinfo
  {year} {2017})}\BibitemShut {NoStop}%
\bibitem [{\citenamefont {Pi}\ and\ \citenamefont {Zocchi}(2021)}]{Ziqi2021}%
  \BibitemOpen
  \bibfield  {author} {\bibinfo {author} {\bibfnamefont {Z.}~\bibnamefont
  {Pi}}\ and\ \bibinfo {author} {\bibfnamefont {G.}~\bibnamefont {Zocchi}},\
  }\href@noop {} {\bibfield  {journal} {\bibinfo  {journal} {J. Phys. Commun.}\
  }\textbf {\bibinfo {volume} {5}},\ \bibinfo {pages} {125013} (\bibinfo {year}
  {2021})}\BibitemShut {NoStop}%
\bibitem [{\citenamefont {Prescott}\ \emph {et~al.}(2008)\citenamefont
  {Prescott}, \citenamefont {Koninck},\ and\ \citenamefont
  {Sejnowski}}]{Sejnowski2008}%
  \BibitemOpen
  \bibfield  {author} {\bibinfo {author} {\bibfnamefont {S.~A.}\ \bibnamefont
  {Prescott}}, \bibinfo {author} {\bibfnamefont {Y.~D.}\ \bibnamefont
  {Koninck}},\ and\ \bibinfo {author} {\bibfnamefont {T.~J.}\ \bibnamefont
  {Sejnowski}},\ }\href@noop {} {\bibfield  {journal} {\bibinfo  {journal}
  {PLoS Comput. Biol.}\ }\textbf {\bibinfo {volume} {4}},\ \bibinfo {pages}
  {e1000198} (\bibinfo {year} {2008})}\BibitemShut {NoStop}%
\bibitem [{\citenamefont {Hodgkin}\ and\ \citenamefont
  {Huxley}(1952)}]{hodgkin_quantitative_1952}%
  \BibitemOpen
  \bibfield  {author} {\bibinfo {author} {\bibfnamefont {A.~L.}\ \bibnamefont
  {Hodgkin}}\ and\ \bibinfo {author} {\bibfnamefont {A.~F.}\ \bibnamefont
  {Huxley}},\ }\href {https://doi.org/10.1113/jphysiol.1952.sp004764}
  {\bibfield  {journal} {\bibinfo  {journal} {J. Physiol. (Lond.)}\ }\textbf
  {\bibinfo {volume} {117}},\ \bibinfo {pages} {500} (\bibinfo {year}
  {1952})}\BibitemShut {NoStop}%
\bibitem [{\citenamefont {Schmidt}\ \emph {et~al.}(2009)\citenamefont
  {Schmidt}, \citenamefont {Cross},\ and\ \citenamefont
  {MacKinnon}}]{MacKinnon_gatingModel2009}%
  \BibitemOpen
  \bibfield  {author} {\bibinfo {author} {\bibfnamefont {D.}~\bibnamefont
  {Schmidt}}, \bibinfo {author} {\bibfnamefont {S.~R.}\ \bibnamefont {Cross}},\
  and\ \bibinfo {author} {\bibfnamefont {R.}~\bibnamefont {MacKinnon}},\ }\href
  {https://doi.org/10.1016/j.jmb.2009.05.062} {\bibfield  {journal} {\bibinfo
  {journal} {Journal of Molecular Biology}\ }\textbf {\bibinfo {volume}
  {390}},\ \bibinfo {pages} {902} (\bibinfo {year} {2009})}\BibitemShut
  {NoStop}%
\bibitem [{\citenamefont {Morris}\ and\ \citenamefont
  {Lecar}(1981)}]{Morris_Lecar}%
  \BibitemOpen
  \bibfield  {author} {\bibinfo {author} {\bibfnamefont {C.}~\bibnamefont
  {Morris}}\ and\ \bibinfo {author} {\bibfnamefont {H.}~\bibnamefont {Lecar}},\
  }\href@noop {} {\bibfield  {journal} {\bibinfo  {journal} {Biophys. J.}\
  }\textbf {\bibinfo {volume} {35}},\ \bibinfo {pages} {193 } (\bibinfo {year}
  {1981})}\BibitemShut {NoStop}%
\bibitem [{\citenamefont {Koch}(1999)}]{Koch_Book}%
  \BibitemOpen
  \bibfield  {author} {\bibinfo {author} {\bibfnamefont {C.}~\bibnamefont
  {Koch}},\ }\href@noop {} {\emph {\bibinfo {title} {Biophysics of
  Computation}}}\ (\bibinfo  {publisher} {Oxford University Press},\ \bibinfo
  {year} {1999})\BibitemShut {NoStop}%
\bibitem [{\citenamefont {Ruta}\ \emph {et~al.}(2003)\citenamefont {Ruta},
  \citenamefont {Jiang}, \citenamefont {Lee}, \citenamefont {Chen},\ and\
  \citenamefont {MacKinnon}}]{ruta_functional_2003}%
  \BibitemOpen
  \bibfield  {author} {\bibinfo {author} {\bibfnamefont {V.}~\bibnamefont
  {Ruta}}, \bibinfo {author} {\bibfnamefont {Y.}~\bibnamefont {Jiang}},
  \bibinfo {author} {\bibfnamefont {A.}~\bibnamefont {Lee}}, \bibinfo {author}
  {\bibfnamefont {J.}~\bibnamefont {Chen}},\ and\ \bibinfo {author}
  {\bibfnamefont {R.}~\bibnamefont {MacKinnon}},\ }\href
  {https://doi.org/10.1038/nature01473} {\bibfield  {journal} {\bibinfo
  {journal} {Nature}\ }\textbf {\bibinfo {volume} {422}},\ \bibinfo {pages}
  {180} (\bibinfo {year} {2003})}\BibitemShut {NoStop}%
\bibitem [{\citenamefont {Qi}\ and\ \citenamefont {Zocchi}(2022)}]{Xinyi2022}%
  \BibitemOpen
  \bibfield  {author} {\bibinfo {author} {\bibfnamefont {X.}~\bibnamefont
  {Qi}}\ and\ \bibinfo {author} {\bibfnamefont {G.}~\bibnamefont {Zocchi}},\
  }\href@noop {} {\bibfield  {journal} {\bibinfo  {journal} {EPL}\ }\textbf
  {\bibinfo {volume} {137}},\ \bibinfo {pages} {12005} (\bibinfo {year}
  {2022})}\BibitemShut {NoStop}%
\bibitem [{\citenamefont {Tsumoto}\ \emph {et~al.}(2006)\citenamefont
  {Tsumoto}, \citenamefont {Kitajima}, \citenamefont {Yoshinaga}, \citenamefont
  {Aihara},\ and\ \citenamefont {Kawakami}}]{BifurcationsMorrisLecar2006}%
  \BibitemOpen
  \bibfield  {author} {\bibinfo {author} {\bibfnamefont {K.}~\bibnamefont
  {Tsumoto}}, \bibinfo {author} {\bibfnamefont {H.}~\bibnamefont {Kitajima}},
  \bibinfo {author} {\bibfnamefont {T.}~\bibnamefont {Yoshinaga}}, \bibinfo
  {author} {\bibfnamefont {K.}~\bibnamefont {Aihara}},\ and\ \bibinfo {author}
  {\bibfnamefont {H.}~\bibnamefont {Kawakami}},\ }\href
  {https://doi.org/10.1016/j.neucom.2005.03.006} {\bibfield  {journal}
  {\bibinfo  {journal} {Neurocomputing}\ }\textbf {\bibinfo {volume} {69}},\
  \bibinfo {pages} {293} (\bibinfo {year} {2006})}\BibitemShut {NoStop}%
\bibitem [{\citenamefont {Liu}\ \emph {et~al.}(2014)\citenamefont {Liu},
  \citenamefont {Liu},\ and\ \citenamefont
  {Liu}}]{BifurcationAnalysisMorris2014}%
  \BibitemOpen
  \bibfield  {author} {\bibinfo {author} {\bibfnamefont {C.}~\bibnamefont
  {Liu}}, \bibinfo {author} {\bibfnamefont {X.}~\bibnamefont {Liu}},\ and\
  \bibinfo {author} {\bibfnamefont {S.}~\bibnamefont {Liu}},\ }\href
  {https://doi.org/10.1007/s00422-013-0580-4} {\bibfield  {journal} {\bibinfo
  {journal} {Biological Cybernetics}\ }\textbf {\bibinfo {volume} {108}},\
  \bibinfo {pages} {75} (\bibinfo {year} {2014})}\BibitemShut {NoStop}%
\bibitem [{\citenamefont {Ariyaratne}\ and\ \citenamefont
  {Zocchi}(2013)}]{Amila2013}%
  \BibitemOpen
  \bibfield  {author} {\bibinfo {author} {\bibfnamefont {A.}~\bibnamefont
  {Ariyaratne}}\ and\ \bibinfo {author} {\bibfnamefont {G.}~\bibnamefont
  {Zocchi}},\ }\href@noop {} {\bibfield  {journal} {\bibinfo  {journal} {PRX}\
  }\textbf {\bibinfo {volume} {3}},\ \bibinfo {pages} {011010} (\bibinfo {year}
  {2013})}\BibitemShut {NoStop}%
\bibitem [{\citenamefont {Strogatz}(2015)}]{Strogatz_Book}%
  \BibitemOpen
  \bibfield  {author} {\bibinfo {author} {\bibfnamefont {S.~H.}\ \bibnamefont
  {Strogatz}},\ }\href@noop {} {\emph {\bibinfo {title} {Nonlinear Dynamics and
  Chaos}}}\ (\bibinfo  {publisher} {Westview Press},\ \bibinfo {year}
  {2015})\BibitemShut {NoStop}%
\bibitem [{\citenamefont {Goldenfeld}(2018)}]{Nigel_Book}%
  \BibitemOpen
  \bibfield  {author} {\bibinfo {author} {\bibfnamefont {N.}~\bibnamefont
  {Goldenfeld}},\ }\href@noop {} {\emph {\bibinfo {title} {Lectures On Phase
  Transitions And The Renormalization Group (Frontiers in Physics v. 85)}}}\
  (\bibinfo  {publisher} {CRC Press},\ \bibinfo {year} {2018})\BibitemShut
  {NoStop}%
\bibitem [{\citenamefont {Hille}(2001)}]{hilleIonChannelsExcitable2001}%
  \BibitemOpen
  \bibfield  {author} {\bibinfo {author} {\bibfnamefont {B.}~\bibnamefont
  {Hille}},\ }\href@noop {} {\emph {\bibinfo {title} {Ion channels of excitable
  membranes}}},\ \bibinfo {edition} {3rd}\ ed.\ (\bibinfo  {publisher}
  {Sinauer},\ \bibinfo {address} {Sunderland, Mass},\ \bibinfo {year}
  {2001})\BibitemShut {NoStop}%
\bibitem [{\citenamefont {Wilders}(2006)}]{DynamicClampPowerful2006}%
  \BibitemOpen
  \bibfield  {author} {\bibinfo {author} {\bibfnamefont {R.}~\bibnamefont
  {Wilders}},\ }\href {https://doi.org/10.1113/jphysiol.2006.115840} {\bibfield
   {journal} {\bibinfo  {journal} {The Journal of Physiology}\ }\textbf
  {\bibinfo {volume} {576}},\ \bibinfo {pages} {349} (\bibinfo {year}
  {2006})}\BibitemShut {NoStop}%
\bibitem [{\citenamefont {Ranjan}\ \emph {et~al.}(2019)\citenamefont {Ranjan},
  \citenamefont {Logette}, \citenamefont {Marani}, \citenamefont {Herzog},
  \citenamefont {T{\^a}che}, \citenamefont {Scantamburlo}, \citenamefont
  {Buchillier},\ and\ \citenamefont {Markram}}]{ranjanKineticMapHomomeric2019}%
  \BibitemOpen
  \bibfield  {author} {\bibinfo {author} {\bibfnamefont {R.}~\bibnamefont
  {Ranjan}}, \bibinfo {author} {\bibfnamefont {E.}~\bibnamefont {Logette}},
  \bibinfo {author} {\bibfnamefont {M.}~\bibnamefont {Marani}}, \bibinfo
  {author} {\bibfnamefont {M.}~\bibnamefont {Herzog}}, \bibinfo {author}
  {\bibfnamefont {V.}~\bibnamefont {T{\^a}che}}, \bibinfo {author}
  {\bibfnamefont {E.}~\bibnamefont {Scantamburlo}}, \bibinfo {author}
  {\bibfnamefont {V.}~\bibnamefont {Buchillier}},\ and\ \bibinfo {author}
  {\bibfnamefont {H.}~\bibnamefont {Markram}},\ }\href
  {https://doi.org/10.3389/fncel.2019.00358} {\bibfield  {journal} {\bibinfo
  {journal} {Frontiers in Cellular Neuroscience}\ }\textbf {\bibinfo {volume}
  {13}},\ \bibinfo {pages} {358} (\bibinfo {year} {2019})}\BibitemShut
  {NoStop}%
\bibitem [{\citenamefont {Jeon}\ \emph {et~al.}(2006)\citenamefont {Jeon},
  \citenamefont {Malmstadt},\ and\ \citenamefont
  {Schmidt}}]{jeonHydrogelEncapsulatedLipidMembranes2006}%
  \BibitemOpen
  \bibfield  {author} {\bibinfo {author} {\bibfnamefont {T.-J.}\ \bibnamefont
  {Jeon}}, \bibinfo {author} {\bibfnamefont {N.}~\bibnamefont {Malmstadt}},\
  and\ \bibinfo {author} {\bibfnamefont {J.~J.}\ \bibnamefont {Schmidt}},\
  }\href {https://doi.org/10.1021/ja056901v} {\bibfield  {journal} {\bibinfo
  {journal} {Journal of the American Chemical Society}\ }\textbf {\bibinfo
  {volume} {128}},\ \bibinfo {pages} {42} (\bibinfo {year} {2006})}\BibitemShut
  {NoStop}%
\bibitem [{\citenamefont {Jeon}\ \emph {et~al.}(2008)\citenamefont {Jeon},
  \citenamefont {Malmstadt}, \citenamefont {Poulos},\ and\ \citenamefont
  {Schmidt}}]{jeonBlackLipidMembranes2008}%
  \BibitemOpen
  \bibfield  {author} {\bibinfo {author} {\bibfnamefont {T.-J.}\ \bibnamefont
  {Jeon}}, \bibinfo {author} {\bibfnamefont {N.}~\bibnamefont {Malmstadt}},
  \bibinfo {author} {\bibfnamefont {J.~L.}\ \bibnamefont {Poulos}},\ and\
  \bibinfo {author} {\bibfnamefont {J.~J.}\ \bibnamefont {Schmidt}},\ }\href
  {https://doi.org/10.1116/1.2948314} {\bibfield  {journal} {\bibinfo
  {journal} {Biointerphases}\ }\textbf {\bibinfo {volume} {3}},\ \bibinfo
  {pages} {FA96} (\bibinfo {year} {2008})}\BibitemShut {NoStop}%
\end{thebibliography}%

\end{document}